\documentclass[12pt]{article}

\usepackage{amssymb}
\usepackage{graphicx}        
\usepackage{makeidx}
\usepackage{epsf}

\makeindex

\def\fnote#1#2{\begingroup\def\thefootnote{#1}\footnote{#2}\addtocounter{footnote}{-1}\endgroup}

\def\inbar{\vrule height1.5ex width.4pt depth0pt}
\def\IB{\relax{\rm I\kern-.18em B}}
\def\IC{\relax\,\hbox{$\inbar\kern-.3em{\rm C}$}}
\def\ID{\relax{\rm I\kern-.18em D}}
\def\IE{\relax{\rm I\kern-.18em E}}
\def\IF{\relax{\rm I\kern-.18em F}}
\def\IG{\relax\,\hbox{$\inbar\kern-.3em{\rm G}$}}
\def\IH{\relax{\rm I\kern-.18em H}}
\def\II{\relax{\rm I\kern-.18em I}}
\def\IK{\relax{\rm I\kern-.18em K}}
\def\IL{\relax{\rm I\kern-.18em L}}
\def\IM{\relax{\rm I\kern-.18em M}}
\def\IN{\relax{\rm I\kern-.18em N}}
\def\IO{\relax\,\hbox{$\inbar\kern-.3em{\rm O}$}}
\def\IP{\relax{\rm I\kern-.18em P}}
\def\IQ{\relax\,\hbox{$\inbar\kern-.3em{\rm Q}$}}
\def\IR{\relax{\rm I\kern-.18em R}}
\def\IT{\relax{\rm I\kern-.18em T}}
\def\ZZ{\relax{\sf Z\kern-.4em Z}}

\def\a{\alpha}   \def\b{\beta}    \def\g{\gamma}  
\def\e{\epsilon} \def\G{\Gamma}     
\def\L{\Lambda}  \def\om{\omega}   \def\si{\sigma}

\def\cA{{\cal A}}      \def\cF{{\cal F}}
 \def\cH{{\cal H}} \def\cI{{\cal I}}   
   \def\cP{{\cal P}}  \def\cR{{\cal R}}
\def\cS{{\cal S}}

\def\Ewhat{{\widehat E}}

\def\adot{{\dot{a}}}

\def\cRdot{{\dot{\cR}}}     

\def\phidot{{\dot{\phi}}}   \def\sidot{{\dot{\si}}}

      \def\rmGeV{{\rm GeV}}

            \def\rmIm{{\rm Im}}     
      
\def\rmMeV{{\rm MeV}}

\def\rmPl{{\rm Pl}}            
       \def\rmSL{{\rm SL}}      \def\rmSO{{\rm SO}}
         
   \def\rmSU{{\rm SU}}

             \def\rmconst{{\rm const}}  
      \def\rmcov{{\rm cov}}         
       \def\rmdet{{\rm det}}             
           \def\rmfs{{\rm fs}}

       \def\rmmin{{\rm min}}     \def\rmmod{{\rm mod}}

          \def\rmvol{{\rm vol}}

\def\rmRe{{\rm Re}}     

  \def\mathR{{\mathbb R}}
\def\mathZ{{\mathbb Z}}

\def\fnote#1#2{\begingroup\def\thefootnote{#1}\footnote{#2}\addtocounter{footnote}{-1}\endgroup}

\def\beq{\begin{equation}}
\def\eeq{\end{equation}}
\def\bea{\begin{eqnarray}}
\def\eea{\end{eqnarray}}
\def\llea#1{\label{#1}\eea}
\def\lleq#1{\label{#1}\eeq}
\let\nn=\nonumber

\def\notin{\ \hbox{{$\in$}\kern-.51em\hbox{/}}}

\def\ra{{\rightarrow}}
\def\lra{\longrightarrow}

\def\del{\partial}

  \def\E1Fq{E_1/\IF_q}

\def\notdiv{{\relax{~|\kern-.34em /~}}}

\def\oE{{\overline{E}}}   \def\oF{{\overline{F}}}

 \def\otau{{\overline{\tau}}}

\def\boxit#1{
\vbox{\hrule height1pt\hbox{\vrule width1pt\kern0.3cm
\vbox{\kern0.3cm\hbox{$\displaystyle#1$}\kern0.3cm}\kern0.3cm\vrule
width1pt}\hrule height1pt}}

\begin{document}

\phantom{\hfill \today}

\parindent=0pt

\vskip .5truein

\centerline{\large {\bf Large and Small Field Inflation from Hyperbolic Sigma Models}}

\vskip .4truein

 \centerline{{\sc~ Rolf Schimmrigk}\fnote{$\diamond$}{netahu@yahoo.com, rschimmr@iusb.edu}}

\vskip .3truein

\parskip 0.2truein

\centerline{Dept. of Physics}

\centerline{Indiana University South Bend} 

\centerline{1700 Mishawaka Ave., South Bend, IN 466344}

\vskip 0.8truein
\parskip=0truein
\baselineskip=16pt

\centerline{\bf Abstract} 

\begin{quote}
  Long standing themes in inflation include the issue of large field vs. small field inflation as well as the
   question what fraction of phase space leads to sufficient inflation, and furthermore is compatible 
   with the experimental data.
   In the present paper these issues are discussed in the context of modular inflation, a specialization of
   the framework of automorphic nonlinear $\si$-models associated to homogeneous spaces $G/K$ in 
   which the continuous shift symmetry group $G$ is weakly broken to discrete subgroups $\G$.
  The target spaces of these theories  inherit a curved structure from the group $G$, which in 
  the case of modular invariant inflation leads to a hyperbolic field space geometry. 
  It is shown that in this class of models the symmetry structure leads to both large and  small field 
  inflationary trajectories within a single modular inflation model. 
  The present paper analyzes the concrete model of $j$-inflation, a hyperbolic model with 
  nontrivial inflaton interactions. It describes  in some detail the structure of the initial conditions, 
  including a systematic analysis of several phenomenological functions on the target space, leading 
  to constraints on the curvature scalar of the field space by upcoming experiments, as well as a discussion of the 
   scaling behavior of the spectral index, the finite volume fraction of the field space leading to sufficient 
   inflation, the attractor behavior of $j$-inflation, and a comparison of inflaton trajectories vs. target space 
   geodesics. The tensor-ratio analysis shows that $j$-inflation is an interesting target for upcoming ground 
   and satellite experiments.
\end{quote}

\renewcommand\thepage{}
\newpage
\parindent=0pt

 \pagenumbering{arabic}

\vskip .1truein

\baselineskip=17.5pt
\parskip=.02truein

\tableofcontents

\vfill \eject

\parskip=.1truein

\section{Introduction} 
 
 The CMB experiments that have been conducted  over the past decade have led to dramatic constraints on the parameters 
  of the models that  make up the extensive landscape of the inflationary  theory space  \cite{planck18.6, planck18.10, a18etal}.
  While most of the progress has come from satellite probes, such experiments are currently  continued on the ground 
   in several observatories around the globe, the results of which will provide further insight into the theoretical parameter 
   space \cite{simons18etal, cmbs4-19a-etal, class21}.  
  While the observables measured or bounded by these experiments provide  constraints strong enough to exclude whole 
 classes of models at high confidence levels, the surviving theory space remains large and unstructured.
 The amorphous nature of this space can be ameliorated somewhat by the introduction of symmetry groups into multifield 
 inflation that are motivated by the fact that they provide an embedding of the shift symmetry. Shift symmetries 
 are often introduced as an ad-hoc device  to ensure the absence of terms that otherwise might affect the small parameters 
 that enter the  observables of the theory. 
 Embedding these symmetries into a proper discrete group has the benefit of allowing
  a quantitative characterization of the  model space by leading to a foliation structure.
  
In the framework of automorphic inflation the  groups that contain the shift symmetry arise from 
nonlinear sigma models associated to homogeneous spaces $G/K$ constructed from reductive groups $G$  
and maximal compact subgroups $K$. The resulting field space is curved with a target space metric 
that is determined by the group $G$ \cite{rs14, rs15}. 
The discrete groups are  obtained  from potentials that weakly break 
the continuous symmetry group $G(\mathR)$  to subgroups  $\G\subset G(\mathZ)$ defined over integers $\mathZ$.  
 In the simplest case of modular inflation the discrete shift symmetry results from the weak breaking of the continuous 
 M\"obius group $\rmSL(2,\mathR)$ to congruence subgroups $\G$ of the modular group $\rmSL(2,\mathZ)$.
 This approach was introduced for the case of  level one modularity in \cite{rs16, rs17} and for higher level in \cite{ls19}. 
The field space of modular inflation carries a nontrivial hyperbolic geometry that derives from the homogeneous 
space $G/K = \rmSL(2,\mathR)/\rmSO(2,\mathR)$ and it  was shown in \cite{rs16} that this metric has important
 implications for  the theory  in that the geometry induces 
 new terms in the observables, as compared to a flat theory, that makes  these observables quasi-modular,
 thereby preserving the modularity, but not the underlying holomorphic nature. This resolves a general issue of 
 modular invariant inflation that was left open  in earlier work in  the particular context of supergravity. 
 The idea to embed the shift symmetry into the modular group and to 
  focus on the hyperbolic geometry was emphasized after the appearance of \cite{rs14}  also in the papers
   \cite{c15etal-a, c15etal-b}.  Recent discussions concerned with the shift symmetry 
  include \cite{a19etal, aw19}.
    
 The concrete models briefly explored in earlier work include the model of $j$-inflation at modular level one \cite{rs14, rs16}
  and the model of $h_2$-inflation at level two \cite{ls19}. It was shown in these papers that these models   
admit inflaton trajectories that are compatible with the {\sc Planck} data.
 The compatibility of $j$-inflation with experiments thus represents  the first hyperbolic twofield inflation  model 
 consistent with the current cosmological data. Work prior to \cite{rs14} based on hyperbolic field space aimed at 
 models with singlefield potentials or effectively singlefield trajectories, for example \cite{kl13, klr14}. Subsequent work 
 with a focus on inflaton orbits often adopted isomorphic versions such as the Poincar\'e disk or the Beltrami-Klein models. 
 This includes for example the references \cite{kl15, a17etal}  and  
  \cite{b17, ls17, bl17, mm17,  l18etal, abl19, m19etal, b20etal, apr20}.
Inflationary models based on negatively curved target spaces have also been considered in other recent work 
\cite{c18etal,  bm19,  c19etal, g19etal, a19etal, w19, kl19} and have played a role in 
the discussion of the conjectures concerned with the  swampland \cite{ls19, m19etal,  b19etal}.

The goal of the present paper is to give a systematic analysis of modular inflation by considering in some detail 
the slow-roll phase space of $j$-inflation, relevant for the observable CMB regime. This analysis leads to a
number of observations that are of relevance beyond the specific model considered here.  One of these concerns the 
issue of large and small field inflation.  The distinction between inflaton fields with values larger or smaller than the 
Planck mass has been a recurrent theme and has often been used as a classification scheme to characterize different types 
of inflationary models.  In modular inflation it is possible to decompose the field space into separate regions 
the union of which gives the total target space. The structures that appear in these separate regions are then repeated 
infinitely many times in the field space.  The nontrivial hyperbolic geometry of these individual regions is given
 by the metric considered originally by Liouville, Riemann, Beltrami and Klein, hence is naturally called 
 the Poincar\'e metric in the literature. This metric leads to a stretching of the field space as one approaches 
  the boundary of the space,  in particular when approaching the origin of the space. 
  This leads to trajectories with initial values that can approach the origin  arbitrarily close, 
  thus leading to initial field values that become small at any scale, in particular compared to the Planck mass. 
Far away from the origin 
one or both of the inflaton components can be large, leading to large field inflation. This shows that in modular 
inflation, and more generally in some curved field space inflation, both small and large field inflation can be realized 
in the same model.

 Modular inflation is a field theory framework with curved target space geometry,
hence it is of interest to ask whether the dynamics of the theory imposes constraints that lead to deviations
 from the geodesic structure induced by the metric.  It will be shown 
 that while in $j$-inflation per se the inflaton trajectories are strongly affected by the potential, they also 
 show attractor behavior in the sense that there are certain basins in the target space to which the trajectories converge.
 
After the discussion of the modular symmetry and the comparison of $j$-inflation trajectories with hyperbolic 
geodesics, the first phenomenological focus of the present exploration will be to identify those regions of the target 
space that lead to  inflation with a number of e-folds in some standard range. This first step is similar in spirit to 
 earlier investigations of other inflationary models such as hybrid inflation
     (see e.g. \cite{l93, l96etal, lv96, t97, ml00, cr08, c09etal, c10, ep13, cr14}) 
  and other models \cite{cmrt18}, which focused 
  on the question of the existence of inflation per se, implemented as a lower bound on the   number of e-folds.
In the context of considering the global scale structure of the field space 
it is of interest that in modular inflation the target space comes equipped with 
a  finite  measure  that allows to quantify the volume in phase space that lead 
to inflation. This is possible without the need of a regularization of this measure, an issue 
 that has afflicted many discussions in this context \cite{ghs86, hp87, gt06, ct10, sw12, rc13}. 

 In the context of sufficient inflation the spectral index and the tensor ratio are left unconstrained.  
 The data obtained by extending  the analysis to these parameters  
  can be used to address a question that concerns the possible scaling of the spectral index and 
 the tensor ratio. This was explored for singlefield inflation in several papers \cite{m13, r13, c14etal}, 
 with the aim of providing a priori constraints in the plane spanned by these two parameters. 
 The important question is what the behavior is  of these two parameters in multifield inflation. 
 Finally, given the experimental data obtained from the CMB satellites it is of interest to consider how
 these observational constraints constrain the regions of viable initial conditions beyond those obtained 
  by the e-fold constraint alone.

 The outline of this paper is as follows.  
 In sections two and three multifield inflation and modular inflation are  briefly reviewed to establish the
           general context and the notation used here. 
 Section four describes the global structure of the potential of $j$-inflation.
 Section five discusses the modular structure of the critical points  on the potential surface and explains the associated 
       existence structure of both large field and small field inflation in the framework of modular invariant sigma models.
  Section six shows that in the global target space there are broad basins of attraction and compares the 
          resulting inflaton trajectories with the geodesics of the hyperbolic geometry. 
  Section seven analyzes the behavior of the e-fold function $N_*$ on the target space and shows that a finite volume fraction 
              of the target space leads to sufficient inflation, where the volume is obtained from the nontrivial 
               geometric measure on  the field space.
  Sections eight and nine consider the spectral index as a function on the target space, including a discussion of the scaling 
  behavior of the spectral index. 
  Section ten analyzes the behavior of the tensor-to-scalar ratio on the field space. It is shown how the bounds of $r$ provided 
        by the CMB probes constrain the curvature scalar of the field space and that $j$-inflation presents an interesting 
        target for the upcoming CMB experiments such as the Simons observatory, as well as LightBird and CMB-S4.  
        This leads to a discussion of large field inflation vs. small values of the tensor ratio, as well as of the  swampland conjectures.     
 Section eleven analyzes how the variation of the mass scale of $j$-inflation impacts the constraints of the CMB on the model.
  Section twelve  presents the conclusions.
 
\vskip .3truein

\section{Formulation of general multifield inflation}

Automorphic inflation is a multifield inflation framework with fields $\phi^I, I=1,...,n$ and a nontrivial target space 
metric $G_{IJ}(\phi^K)$ that leads to a nontrivial kinetic coupling
action
\beq
 \cA ~=~ - \int d^4x \sqrt{-g} \left(\frac{1}{2} G_{IJ}(\phi^K) g^{\mu \nu} \del_\mu \phi^I \del_\nu\phi^J ~+~ V(\phi^I)\right).
\eeq
 The perturbations relevant in this general context 
can be parametrized as $(\cR, S^{IJ})$, where $\cR$ is the 
comoving curvature perturbation of Lukash and Bardeen \cite{l80, b80} (see also \cite{w08})
\beq
 \cR ~=~ H\delta u ~-~ \psi,
\eeq
in terms of the spacetime metric perturbation $\psi$, 
 and $S^{IJ}$ is the tensor of isocurvature perturbations \cite{rs16}
 \beq
  S^{IJ} ~=~ \frac{H}{\sidot} \left(\si^I Q^J ~-~ \si^J Q^I\right).
 \eeq
Here $\sidot = (G_{IJ}\phidot^I \phidot^J)^{1/2}$ is the speed of the background inflaton  and $\si^I=\phidot^I/\sidot$ is
 the normalized inflaton velocity. The closed form of the dynamics of 
 the system $(\cR,S^{IJ})$ was derived in \cite{rs16} as 
 \bea
 \cRdot &=& - 2H \eta_{IK} \si^K \si_J S^{IJ}  \nn \\
 D_t S^{IJ} &=& 2H\left(\eta_{KL} \si^K\si^L - \e\right) S^{IJ} + H\left(\eta_{KL} G^{K[I} S^{J]L} - 
      \frac{\e}{3} M_\rmPl^2 \si^{[I}R^{J]}_{KLM} \si^K S^{LM}\right),
 \eea
 where 
 \beq
  \eta_{IJ} ~=~ M_\rmPl^2 \frac{V_{;IJ}}{V}
 \eeq
 and $V^{[I}W^{J]} = V^I W^J - V^JW^I$. The covariant derivative $D_t$ acts on the contravariant 
 tensor $S^{IJ}$.
The specialization to the twofield case $(\cR,\cS)$ is discussed in detail in \cite{rs16}.

The parameters considered in the present analysis  include the spectral index $n_{\cR\cR}$ of the scalar power spectrum $\cP_{\cR\cR}$ 
of $\cR$, the tensor ratio  and the number of e-folds
\beq
   \cP_{\cR\cR} ~=~ A_{\cR\cR} \left(\frac{k}{k_*}\right)^{n_{\cR\cR}-1}, ~~~~
   \cP_T ~=~ A_T \left(\frac{k}{k_*}\right)^{n_T}, ~~~~
    r~=~ \frac{\cP_T}{\cP_{\cR\cR}},  ~~~~~~~
   N_*=\int_{t_*}^{t_e} H dt.
\lleq{params-def}
 The general expressions in multifield inflation with an inflaton multiplet $\phi^I$ and a curved target space metric $G_{IJ}$ 
 are given by
\bea
 n_{\cR\cR} &=& 1 - 3G^{IJ}\e_I\e_J ~+~ 2 \frac{\eta_{IJ}\e^I\e^J}{G^{KL}\e_K\e_L} \nn \\
  r &=& 8 G^{IJ}\e_I\e_J \nn \\
  n_T &=& -\frac{r}{8}
 \llea{nRR}
 where the slow-roll parameters $\e_I$ are defined as
 \beq
  \e_I ~:=~ M_\rmPl \frac{V_{,I}}{V},
 \eeq
and $H=\adot/a$ is the Hubble-Slipher parameter.   The slow-roll dynamics 
 can be written in terms of the potential as
  \beq
   \phidot^I ~=~ - \sqrt{\frac{V}{3}} G^{IJ}\e_J 
  \eeq
  
  A systematic analysis of models in such a framework involves in the first instance the variation of the energy scales 
  that define the model, as well as a scan of the phase space of the theory. In the context of the phase space 
  analysis the question of a measure arises, a problem that involves the geometry of the 
  target space. Other questions include the dichotomy  of  large vs. small field inflation, an issue that also  turns out to 
  depend on the geometry of the field space. Finally one can ask whether the  scaling behavior of the CMB observables
  observed in simple singlefield models has some counter part in the multifield context.

\vskip .3truein

\section{Structure  of modular inflation on hyperbolic field space}

This section briefly outlines the essential structures that will be used in the remainder of this paper.
Modular inflation is a framework of twofield inflation that specializes 
multifield theory of automorphic inflation \cite{rs14, rs15}. The latter  is based on automorphic functions 
$F(\phi^I)$, where the inflaton multiplet defines  the coordinates of quotient spaces $G/K$, defined in terms of 
reductive groups $G$ and maximal compact subgroups $K$.

 In the simplest case the doublet  spans the curved target space geometry given by the upper halfplane $\cH$, which 
can be identified as the homogeneous space 
\beq
 \cH ~:=~ \rmSL(2,\mathR)/\rmSO(2,\mathR).
 \eeq
The geometry of this space is determined by  the hyperbolic metric
\beq
 G_{IJ} ~:=~ \frac{\mu^2}{(\phi^2)^2} \delta_{IJ}, ~~~I,J=1,2,
\lleq{hyperbolic-metric}
where $\mu$ is an a priori arbitrary energy scale. This metric of the upper halfplane leads to a space of constant curvature
 \beq
  R ~=~ - \frac{2}{\mu^2}.
 \lleq{field-space-curvature}
 Hence the energy scale $\mu$ can be thought of as a measure of the curvature of the target space. In distinction to the energy scale 
 $\L$, the parameter $\mu$ enters  all the formulae that characterize $j$-inflation. As a result the analyses below 
 will show that the curvature scalar $R$ is constrained by the CMB experiments.
 The hyperbolic modular inflation metric admits an isometry group that contains the M\"obius group and the 
 transformation $\tau \mapsto (-\otau)$. 
  
 In  modular inflation it is convenient to complexify the inflaton doublet as $\phi= \phi^1+i\phi^2$ 
 and to introduce the dimensionless inflaton as $\tau=\phi/\mu$. This recovers the so-called Poincar\'e metric of 
 Liouville, Beltrami and Klein as  $ds^2=\mu^2 d\tau d\otau/(\rmIm~\tau)^2$.  
 An important feature of hyperbolic geometry is that, while the curvature is constant, the metric diverges as it approaches
 the boundary of the space $\phi^2=0$, hence the target space in this parametrization is the upper halfplane  $\phi^2>0$.
 The form of the boundary can be transformed into different shapes if different variables are used, such as in the 
  Poincare disk model which is  obtained via the Cayley transform and which 
was considered later in  \cite{c15etal-a, c15etal-b}.
 Other models include the sinh-form of the metric considered more recently in \cite{b17, mm17, c19etal, b20etal}.
 These related models   of the hyperbolic geometry can be described in terms of homogeneous spaces 
 based on quotients of the groups $\rmSU(1,1)$ and $\rmSO(2,1)$ instead of $\rmSL(2,\mathR)$.
  This set-up can be compared to the metric sometimes considered in
  axion-dilaton inflation with a radial field $\rho$ and an angular field $\theta$ with a semi-flat metric that 
 depends on the radial field only.  Writing the complex field as $\phi = \rho e^{i\theta}$ leads 
   to the modular inflation metric $G_{IJ} = \mu^2\delta_{IJ}/\rho^2 \sin^2\theta$
 involving both target space directions.

Given a hyperbolic metric on the target space with an associated symmetry group it is natural to consider potentials that 
preserve at least some of these symmetries. In modular inflation the idea is to break the continuous M\"obius group 
$\rmSL(2,\mathR)$  to an  infinite discrete group $\G$ of the modular group $\rmSL(2,\mathZ)$ that contains the shift symmetry
\beq
 \rmSL(2, \mathR) ~\stackrel{V_\G}{\lra} ~ \G \subset \rmSL(2,\mathZ)
 \eeq
 where the potential 
  \beq
   V_\G ~=~ \L^4 ~|F|^2
  \eeq
  is defined in terms of a function $F$ that is invariant under the group $\G$. 
  This allows for nontrivial potentials in which both components of 
the inflaton doublet  are coupled in a nontrivial way.  Such modular invariant inflation potentials can be 
constructed in different ways, but the most convenient form for the 
 phenomenological analysis is in terms of the Eisenstein series $E_4, E_6$,
 of weight four and six respectively,  which are defined as
 \beq
  E_w(\tau) ~=~ 1 - \frac{2w}{B_w}\sum_{n\geq 1} \si_{w-1}(n)q^n,
  \lleq{eisenstein-series}
  where $q=e^{2\pi i \tau}$, the denominators $B_w$ are the Bernoulli numbers and $\si_w(n)$ is the divisor function
  \cite{rs14, rs16}.

In general modular inflation the models involve a modular invariant function $F(\phi^I/\mu)$,   hence depend in the 
formulation given here on the energy scale $\mu$  that enters the potential. As a result the slow-roll parameters $\e_I$ 
depend on $\mu$ and can be expressed in the form \cite{rs16} 
\beq
 \e_I ~=~ i^{I-1} \frac{M_\rmPl}{\mu} \left(\frac{F'}{F} + (-1)^{I-1} \frac{\oF'}{\oF}\right),
 \eeq
 leading to
 \beq
 \e_V ~=~ \frac{1}{2} G^{IJ}\e_I\e_J ~=~ 2\frac{M_\rmPl^2}{\mu^2} (\rmIm~\tau)^2 \left|\frac{F'}{F}\right|^2.
 \eeq
 
 The cosmological parameters considered above can be shown to  be expressed in a quasi-modular form as \cite{rs16} 
\bea
 n_{\cR\cR} &=& 1 - 4\frac{M_\rmPl^2}{\mu^2} (\rmIm~\tau)^2 \left[ 2\left|\frac{F'}{F}\right|^2 - 
      \rmRe\left(\frac{F''}{F'} \frac{\oF'}{\oF}\right)_\rmmod + \frac{\pi}{3} \rmIm\left(\Ewhat_2 \frac{\oF'}{\oF}\right)\right]  \nn \\
  r &=& 32 \frac{M_\rmPl^2}{\mu^2} (\rmIm~\tau)^2~\left|\frac{F'}{F}\right|^2 \nn \\
 N_* &=& \frac{1}{\sqrt{3}} \frac{\L^2}{M_\rmPl} \int_{t_*}^{t_e}  dt~ |F(\tau(t)|, 
 \llea{nRRrN-MI}
 where the quasi-modular form $\Ewhat_2$ is defined as
 \beq
  \Ewhat_2(\tau) ~=~ E_2(\tau) - \frac{3}{\pi (\rmIm~\tau)}.
 \eeq
 This shows that the spectral index is a modular invariant quantity, an issue that was not addressed in the 
 earlier literature, for  example in the context of supergravity theories.
  
The model of $j$-inflation is defined at level $N=1$, i.e. it is invariant under the full modular group $\rmSL(2,\mathZ)$
 with $F(\tau) = j(\tau)$ . Here the absolute modular invariant 
 $j(\tau)$-function is viewed as a function of the dimensionless parameter $\tau=\phi/\mu$ in the complex upper 
 halfplane  $\cH$.  The  function $j(\tau)$ is  invariant under the full modular group $\rmSL(2,\mathZ)$,
 a group that contains the shift symmetry of the inflaton. 
  The $j$-function is most naturally defined as the   quotient of two modular forms of weight twelve
    \beq 
  j(\tau) ~:=~ \frac{E_4^3(\tau)}{\Delta(\tau)},
 \eeq
 where $\Delta(\tau) = \eta^{24}(\tau)$ is the Ramanujan modular cusp form of the full modular group, expressed 
 in terms of the Dedekind eta function 
 \beq
  \eta(\tau) ~=~ q^{1/24} \prod_{n\geq 1} (1-q^n).
 \eeq
 The Ramanujan form can be written in terms of the generators  $E_4, E_6$ of the space of all modular forms 
 relative to the modular group $\rmSL(2,\mathZ)$ as $\Delta = (E_4^3 - E_6^2)/1728$. 

 The slow-roll parameters $\e_I$ can now be expressed directly in terms of undifferentiated modular forms as
 \beq
  \e_I ~=~ -2\pi i^I \frac{M_\rmPl}{\mu} \left(\frac{E_6}{E_4} + (-1)^I \frac{\oE_6}{\oE_4}\right) 
  \eeq
 with
 \beq
 \e_V ~=~ 8\pi^2 \frac{M_\rmPl^2}{\mu^2} (\rmIm~\tau)^2 \left|\frac{E_6}{E_4}\right|^2,
 \eeq
  and the above parameters are given directly in terms of the Eisenstein series as  \cite{rs14, rs15}
 \bea
 n_{\cR\cR} &=& 1 -  \frac{8\pi^2}{3} \frac{M_\rmPl^2}{\mu^2} (\rmIm~\tau)^2\left[  8 \left|\frac{E_6}{E_4 }\right|^2 
     - 3 \rmRe\left(\frac{E_4^2}{E_6} \frac{\oE_6}{\oE_4}\right) ~+~ \rmRe \left(\Ewhat_2 \frac{\oE_6}{\oE_4}\right)\right] 
    \nn\\
 r & =  & 128 \pi^2 \frac{M_\rmPl^2}{\mu^2} ~(\rmIm ~\tau)^2  \left|\frac{E_6}{E_4}\right|^2 \nn \\
 N_* &=& \frac{\L^2}{\sqrt{3}M_\rmPl}  \int dt~ |j|.
 \llea{jinfl-params}
  The slow-roll dynamics takes for $j-$inflation the form
 \beq
  \phidot^I ~=~ \frac{2\pi i^I}{\sqrt{3}} \frac{M_\rmPl}{\mu} \L^2 \left(\rmIm~\tau\right)^2 
    \left( \frac{E_6}{E_4} + (-1)^I\frac{\oE_6}{\oE_4}\right)~|j|.
  \eeq
 The energy scale $\L$ does not enter $n_{\cR\cR}$ or $r$, but it does appear in the differential equation and 
 hence the number of e-folds associated to inflaton trajectories. It can be expressed in terms of $\mu$ and the amplitude
  of the primordial CMB amplitude $A_{\cR\cR}$ as
  \beq
   \L ~=~ M_\rmPl \left(192\pi^4 \frac{M_\rmPl^2}{\mu^2} A_{\cR\cR} (\rmIm~\tau)^2 \left|\frac{E_6}{E_4}\right|^2 
    \frac{1}{|j|^2}\right)^{1/4}.
   \eeq
   This leaves the parameter $\mu$, and hence the curvature scalar, to be constrained by the cosmological observables.
The phenomenological analysis now aims to investigate the target space of $j$-inflation. 
 
  A more detailed foundational formulation of modular inflation at level one can be found in \cite{rs14,rs16}, and  the 
  generalization from the full modular group $\rmSL(2,\mathZ)= \G_0(1)$ to Hecke congruence subgroups 
  $\G_0(N)$ at higher level $N$ has been introduced in ref. \cite{ls19}.

\vskip .3truein

\section{Global structure of the potential surface of $j-$inflation}

The aim of the preliminary phenomenological analysis of $j$-inflation presented in ref. \cite{rs16} was to establish the existence of 
inflaton trajectories  that are compatible with the known constraints for the spectral index of the scalar power spectrum,
the bound on the tensor power spectrum via the tensor ratio, and the number of e-folds. 
Since the bound on the  tensor ratio $r$ established by the CMB satellite probes is small, it is clear from the 
$j$-inflation expression for $r$ given in eq. (\ref{jinfl-params}) that viable values for $\phi_*^I$ can be found in the  neighborhoods
$U(\tau_s)$ of the zeros of the modular form $E_6$. One such zero is at $\tau_s=i$.  A close-up view 
of the neighborhood $U(\tau_s)$ for $\tau_s=i$,
 together with the shape of a typical trajectory, is shown in figure 4 of ref. \cite{rs16}. 
 The scale of the saddle point at $\tau_s=i$ is given by the curvature 
matrix, which for the $j$-inflation potential is given by 
\beq
 V_{,IJ}(i) ~=~ \frac{3\cdot 432^2 \G\left(\frac{1}{4}\right)^8}{\pi^4} \frac{\L^4}{\mu^2} \left(\matrix{-1 &0\cr 0 &1\cr}\right).
 \eeq
 This curvature matrix immediately gives the mass matrix of the inflaton perturbations at the critical point. Away from the 
 saddle point the mass matrix receives contributions from terms that depend on the slow-roll parameters. 
 A recent discussion that aims at a characterization of critical points in terms of discrete symmetries in a general context that includes inflationary 
 models can be found in \cite{k20}.
 
 Attempts have been made  to classify singlefield inflation models by the curvature $V''$ of the 
 potential \cite{d97etal, k98, h00etal, p03-wmap1, k03etal}. A natural way to generalize these early discussions
 would be by keeping track of the eigenvalue structure of the flat curvature matrix $V_{,IJ}$ of the potential, or its covariant form $V_{;IJ}$.
 A simpler structure could alternatively be obtained by considering the determinant $J(V) = \rmdet (V_{,IJ})$, 
 or its covariant form $J_\rmcov(V)$. 
 A more phenomenological, but somewhat less structural, characterization however might proceed by the generalization 
  of the singlefield slow-roll parameter $ M_\rmPl^2 V''/V$. In multifield inflation the analog of this parameter is
   $\eta_{IJ}\e^I\e^J/(G^{KL}\e_K\e_L)$, which is not 
  just a rescaled curvature matrix.
  
In the present paper the phenomenological considerations aim at a more global view of the slow-roll phase space and 
the structure of the potential surface away from the saddle point $\tau_s=i$ becomes of interest.
One of the main motivations for $j$-inflation is the fact that the shift symmetry, often introduced as an ad hoc approximate invariance, 
is now part of a bona-fide symmetry of the theory. Modular inflation at level one is invariant under the full modular group 
$\rmSL(2,\mathZ)$ of integral  $2\times 2$ matrices of determinant one.
This infinite group can be generated by just two elements, given by 
 \beq
  S ~=~ \left(\matrix{0 &1 \cr -1 &0\cr}\right), ~~~~T ~=~ \left(\matrix{1 &1\cr 0 &1\cr}\right).
 \lleq{mod-group-gens}
The generator $T$ shifts the inflaton multiplet, leading to a repetitive pattern of the inflaton field space of the 
upper halfplane $\tau\in \cH$ as $T$ shifts by one unit $\tau \lra \tau+1$.  
The investigation of the behavior of modular inflation models can therefore 
be restricted without loss of generality to bands of width one. 
In this paper the focus will be on the vertical band bounded by  $-1/2 \leq \tau^1 < 1/2$.

The structure of the potential surface is simple for larger $\tau^2$, but becomes more complicated 
as one approaches the region where $\tau^2~\ra~ 0$.  Figure 1 illustrates this with a view that shows a partial region 
 of the embedded surface. This graph indicates a canyon like structure that emerges as one approaches
 the boundary of the field space given by the horizontal axis for $(\rmIm ~\tau)=0$, with saddle points visible at $\tau_s=i$
 and $\tau_s=(1+i)/2$. Further saddle points exist closer to the boundary, but are obscured in this 3D picture. 
 These will become transparent in the contour graphs discussed further below.  In coordinate space the slopes steepen 
 as $\rmIm~ \tau~\ra~0$, but the hyperbolic metric  stretches the physical distances in field space, thereby  flattening the 
  potential surface.
   This will become important below in the analysis of the behavior of the trajectories in $j$-inflation.
   \begin{center}
 \includegraphics[scale=0.1]{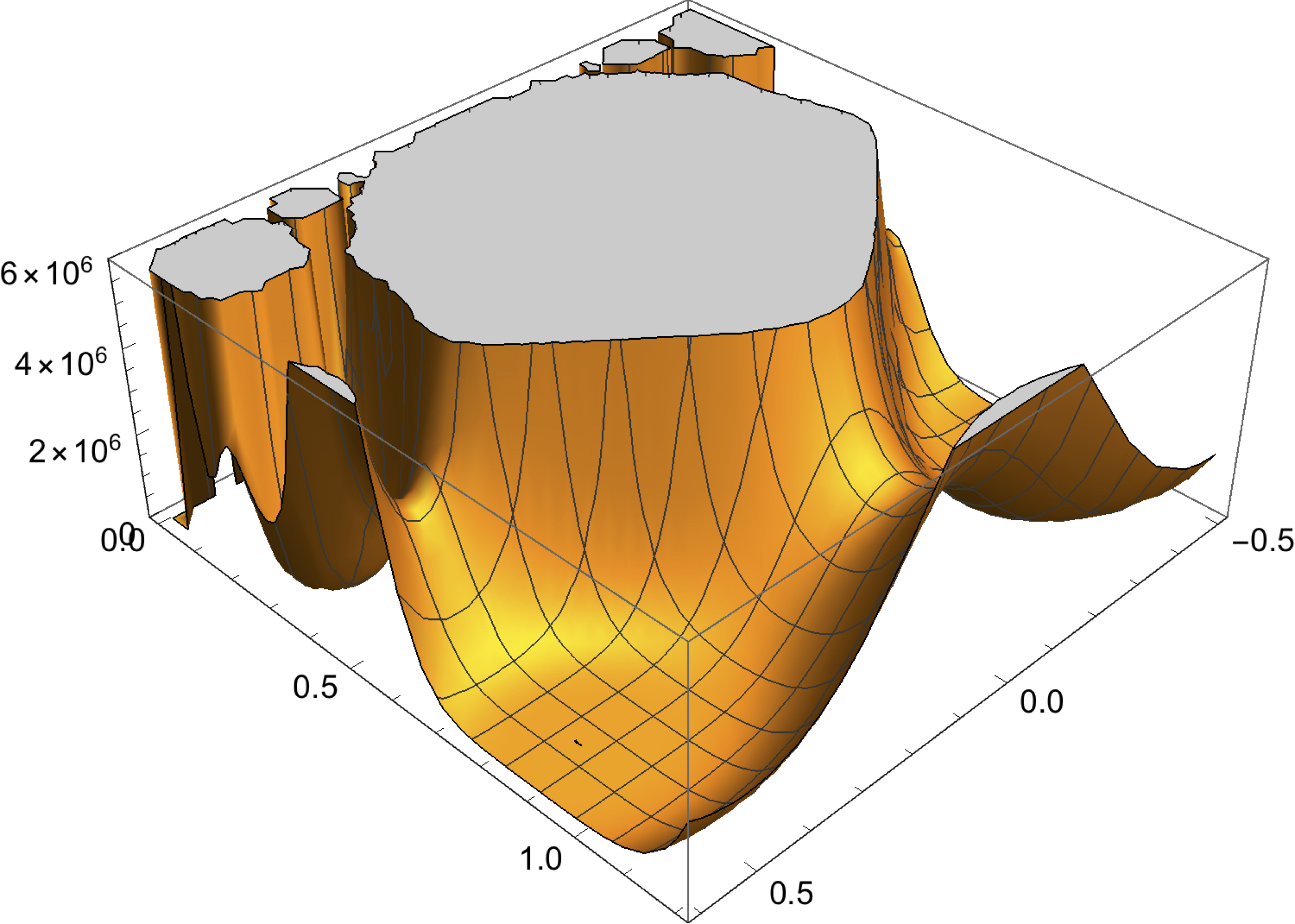}
 \end{center}
\begin{quote}
\baselineskip=15pt
{\bf Figure 1.} ~A large scale view of the $j$-function surface $|j|^2$ (the energy scale parameter $\L$  is 
determined by the CMB amplitude $A_{\cR\cR}$). 
    The plot shows two of the saddle point regions, one of which has a mirror on the far side of the graph.
   The minimum of the potential is located at $\tau_\rmmin = e^{2\pi i/6}$.
   \end{quote}

\baselineskip=17.5pt

 The minimum of the potential at $\tau_\rmmin = e^{2\pi i/6}$, which is determined by the zero of the Eisenstein series $E_4$
  in the numerator of the inflaton potential, combined with the fact that at the point $\tau_\rmmin$ the Ramanujan modular 
  form $\Delta(\tau)$ that appears in the denominator of $V$ does not vanish. 
 As the component $\phi^2$  of the inflaton doublet increases beyond the saddle point at $\tau_s=i$ the $j$-inflation 
  potential forms a wall and diverges. This can be seen explicitly from the definition of the Eisenstein series given in 
  eq. (\ref{eisenstein-series}). As   $\rmIm~\tau~\ra~\infty$ the variables $q=e^{2\pi i \tau}$ vanish and the Eisenstein series
  approach unity in the normalization adopted here. Since the Ramanujan form approaches $q$ in this limit the $j$-function 
  diverges. 

The trajectories of the inflaton evolution considered in \cite{rs16} all originated in the neighborhood 
$U(\tau_s)$ of the saddle point at $\tau_s=i$. The evolution 
of those orbits was such that it lasts long enough to produce sufficient inflation and is furthermore compatible with the observational 
results of the CMB probes.  Figure 1 shows that there are other saddle point regions and
 that the potential surface of $j$-inflation has an intricate structure that the large scale three-dimensional representation 
of the surface does not display in full detail.  Better insight into the potential surface can be obtained by combining the 3D 
potential shown in figure 1 with the contour plot in figure 2. This graph shows the saddle point region considered in \cite{rs16}, 
as well as the structures in figure 1, and also shows a number of further saddle points on both sides of the ridge of the 
saddle point $\tau_s=i$.
A zoom of the above contour plot in the right panel of figure 2 
shows that the potential becomes more complicated as one approaches the horizontal axis, 
the boundary of the target space. This structure is reminiscent of a fractal structure, but the picture suggested by this graph 
has to be complemented by the behavior of the hyperbolic metric, which diverges as the saddles approach the real axis.

\begin{center}
 \includegraphics[scale=0.08]{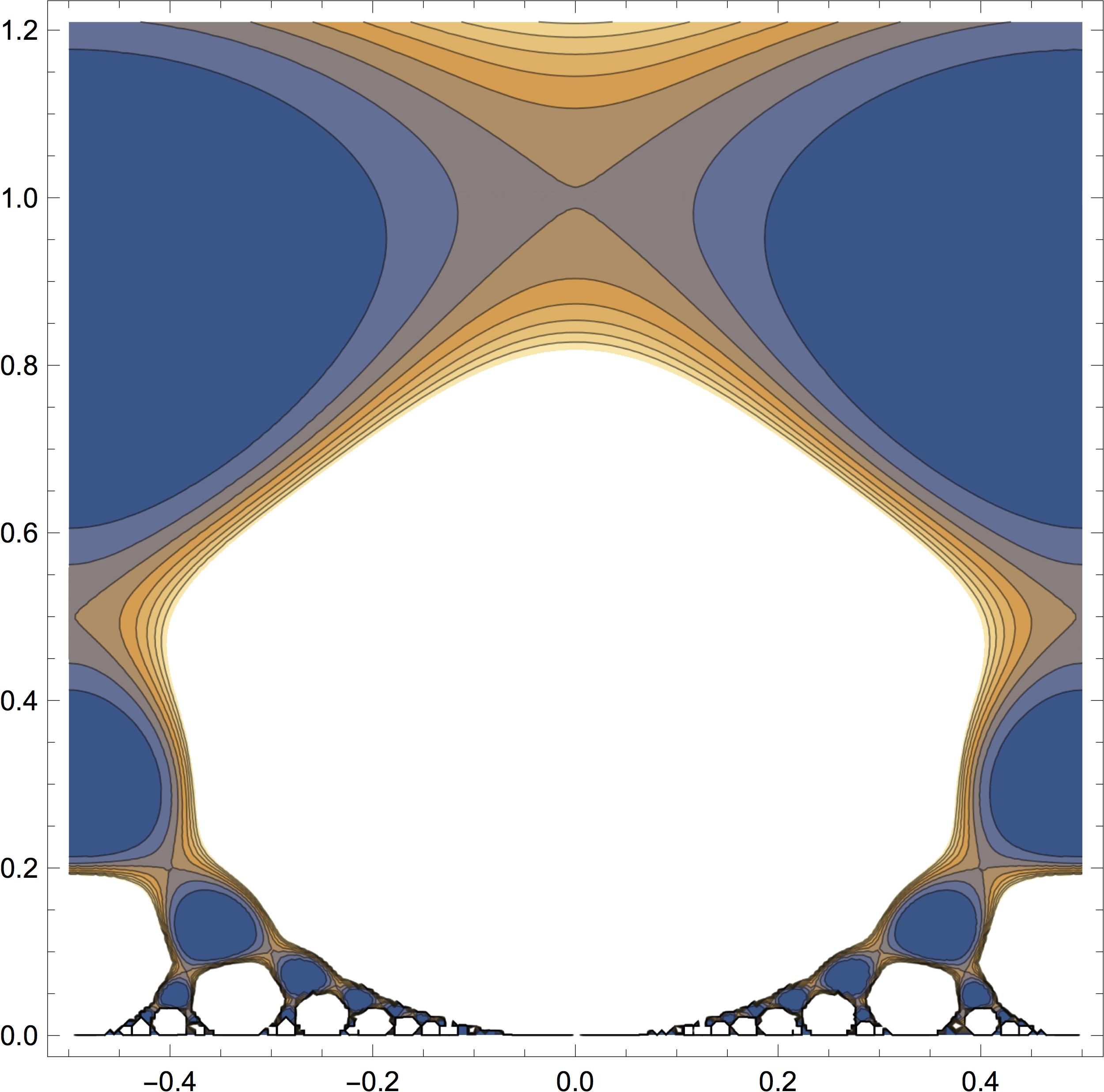}
 ~~~
 \includegraphics[scale=0.08]{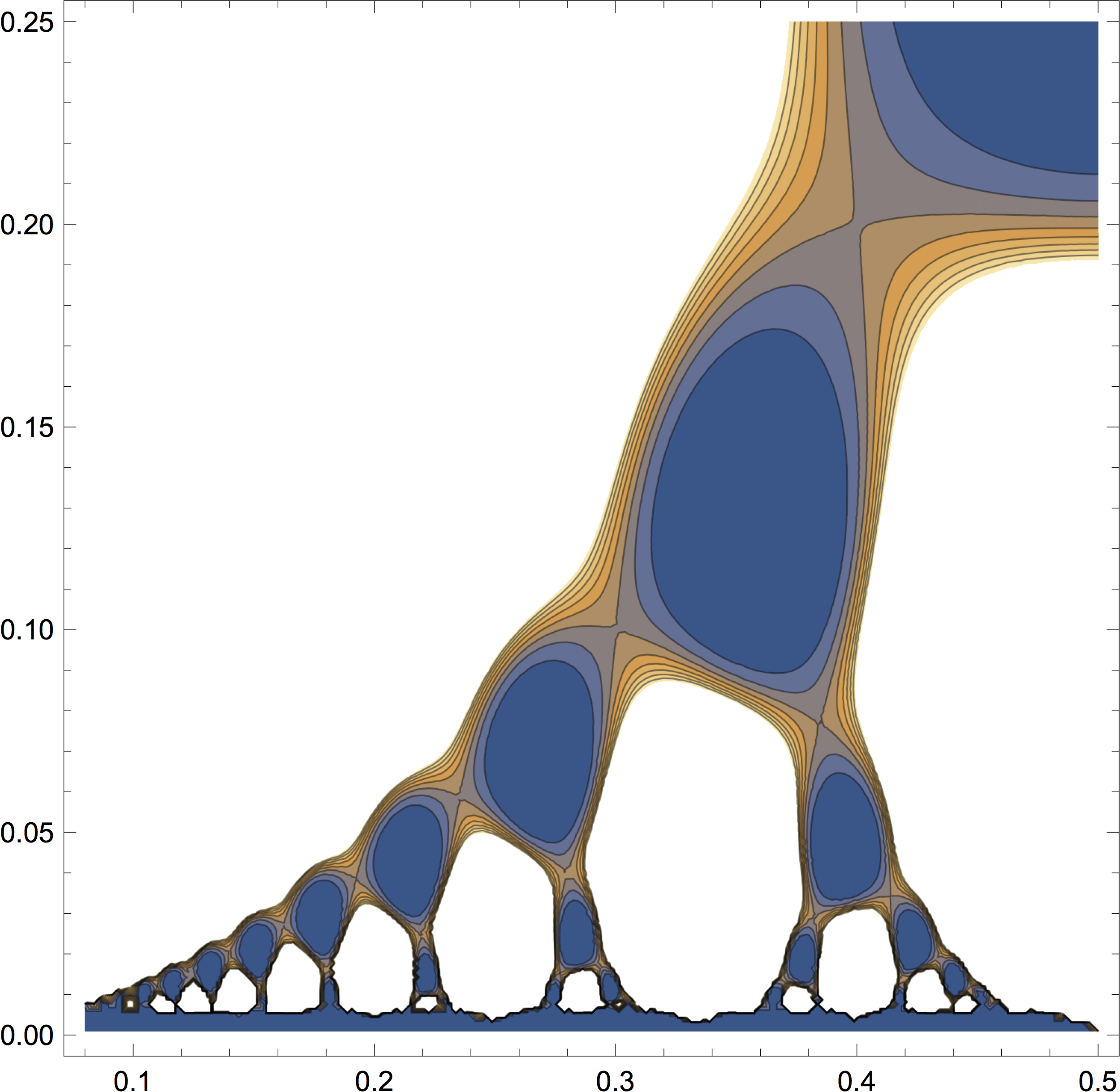}
  \end{center}
\begin{quote}
\baselineskip=15pt
{\bf Figure 2.} ~Contour lines of the potential surface modulo the energy scale. The higher the potential region, the lighter the color.
 This graph shows that there are further saddle point regions when $\tau^2$ approaches the boundary of the target space, 
 determined by $\tau^2=0$.
The ridge toward the top of the left panel shows the region around the saddle point $\tau_s=i$ 
which was the main focus of ref. \cite{rs16}.
\end{quote}

\baselineskip=17.5pt

 The precise degree of detail obtained in the plots of the $j$-inflation potential depends on the resolution and also on the 
 accuracy with which the potential is computed. This is relevant because the modular forms involved in the  
 definitions of the $j$-function are infinite series which have to be truncated.  An increase of the number of terms increases 
 the amount of detail that can be seen, thus the order to which the Eisenstein series are computed 
 needs to be adjusted depending on  how close to the real axis the computation 
 should proceed.

\vskip .3truein
 
 \section{Large and small field inflation in hyperbolic $\si$-models}
  
One of the longstanding issues in inflation is the dichotomy between large field 
and small field inflation, emphasized for example by Lyth in the  context of singlefield inflation \cite{l96}. 
These discussions revolve  around the relation between the distance covered by the 
 inflaton in field space and the tensor ratio, mediated by the number of e-folds during inflation. In multifield 
 inflation this relation is less direct because the distance in field space involves not only changes of the fields $\Delta \phi^I$ 
 but also depends on the metric $G_{IJ}$ of the target space. It is therefore useful to separate issues pertaining to the 
 field values from those that depend on the distance in field space, hence in general involve the metric.
 
 Part of the current discussion that informs the planning and construction of the next generation experiments aiming to 
 discover the 
 contribution of inflationary tensor modes  to the CMB is the formulation of significant targets for the tensor ratio $r$, 
 which will be discussed in detail in section ten.
 Tensor mode targets that have been identified in recent years have been motivated by distinguishing between models with 
 inflaton values that 
 are either super- or sub-Planckian.  This distinction has in particular led to the formulation of classes of singlefield inflation models 
 along these lines in the hope that this leads to the identification of  more generic characteristics of inflationary models that might 
 be testable with CMB data (early work in this direction can be found in the refs.  \cite{d97etal, k98, h00etal, p03-wmap1, k03etal}). 
 The demarcation line between these two classes  is not precise, but this lack of precision is not important for the following. 
 What is of interest is the common association in the literature of large field inflation with  large $r$ and small field inflation 
 with smaller $r$. These discussions raise the question  whether a similar behavior can be identified in multifield inflation  in general.
 
 In $j$-inflation there are two aspects that arise in this context. The first, of relevance in principle for any multifield framework, 
 is that  the different components of the inflaton multiplet can have rather different scales, making the distinction of 
 sub-Planckian vs. super-Planckian field values less a feature of the field itself, even though it still retains its importance 
 for at least some components. This is the case in $j$-inflation in 
 neighborhoods $U(\tau_s)$ of the saddle point $\tau_s=i$.
 For initial values $\tau_*$ in $U(i)$ at least  one of the field values is  super-Planckian for $\mu > M_\rmPl$. 
 Hence the vertical component is large $\phi_*^2 \cong \mu$, while the horizontal component $\phi_*^1$ can be 
  much smaller, depending on how close to the saddle point $\phi_s$ the initial value $\phi_*$ is. 
  In this case the general philosophy can be applied by a slight change of formulation. 
  
  A second feature, specific to the modular inflation framework,  is that modular transformations connect an infinite number of 
  different regions of the target space. This also applies more generally to automorphic inflation. 
  The contour plots in figure 2 show sequences of critical points of the $j$-inflation potential 
  that form   arcs oriented toward the origin of the $\tau$-plane.  These saddle point regions can be mapped with elements of the 
  modular group $\rmSL(2,\mathZ$), which is generated by the elements $S$ and $T$ in eq. (\ref{mod-group-gens}).
  The top arcs in the left panel of figure 2 can be reached for example from the saddle point at $\tau_s=i$ with
  the sequences of group elements
 \beq
  (TST)^{\pm n}(i) ~=~ \pm \frac{n}{n^2+1} + \frac{i}{n^2+1}.
 \lleq{modular-sequence}
 These representations of the primary 
 arcs are not unique because $Si=i$, hence these primary sequences
 can be obtained in a number of different group elements. 
 
 The next sequences of saddle points, both illustrated  in the left panel, but more clearly visible in the zoom in the 
 right panel, can be obtained via 
  \bea
  (TST)^nT(i) &=& \frac{2n+1}{2n^2+2n+1} ~+~ \frac{i}{2n^2+2n+1}  \nn \\
  ST^n(TST)(i) &=& - \frac{2n+1}{2n^2+2n+1} ~+~ \frac{i}{2n^2+2n+1}. 
  \llea{secondary-sequences}
The general picture illustrated by these maps is that modular invariance of the $j$-potential leads to saddle points 
$\tau_n$ obtained by maps $\g_n$ obtained from elements in the modular group.

 The main feature of these maps $\g_n$ is  that they approach the origin $\tau=0$ ever closer as $n$ increases, which is 
 apparent for the images of the $\tau_s$ in (\ref{modular-sequence}) and (\ref{secondary-sequences}). For more general $\tau$
  this can be illustrated with the sequences $\g_n$ obtained for example by the iteration of a generating element 
  $\g$ as $\g_n = \g^n$ for the group element $\g=TST$. These maps $\g_n$
  send arbitrary points $\tau$ to images of the form
 \beq
 \g_n(\tau) ~=~ \frac{\tau^1 + n |\tau|^2 + i\tau^2}{1+2n\tau^1+ n^2|\tau|^2},
 \eeq
  generalizing the sequence (\ref{modular-sequence}) of the saddle point $\tau_s=i$.
 For an  initial value $\tau_*$ in a neighborhood of $ \tau_s=i$, this sequence  converges to the origin of the $(\tau^1,\tau^2)$-plane 
 and in the process sends both components $\tau^I_*$ to ever smaller field values.  
  The existence of such sequences that  approach the origin  of the field space shows that it is possible to have viable $j-$inflation
  trajectories that   start with sub-Planckian field values.

The implication of these sequences of maps $\g_n$ is that large field initial values  $\tau_*$ in a neighborhood of the large 
field saddle point $\tau_s=i$  with larger tensor ratios $r(\tau_*)\geq 0.01$ (which satisfies the {\sc Planck} bound) are mapped 
into small field initial values $\g_n(\tau_*)$  with the same tensor ratio 
 \beq
  r(\g_n(\tau_*)) ~=~ r(\tau_*).
 \eeq
In section ten it will be shown that, depending on the parameters of the model, $j$-inflation can lead to tensor 
ratios that reach the current bound reported by the {\sc Planck} collaboration, but can also reach down to the target range 
of upcoming experiments, such as the Simons Observatory and the CMB-S4 experiment. 
Combined with the discussion above this shows that in $j$-inflation, and eo ipso  in multifield inflation, 
small field inflation can lead to large tensor ratios and large field inflation can 
 lead to small tensor ratios.
 
\vskip .3truein

\section{$j$-Inflation trajectories and hyperbolic geodesics} 

The analysis discussed in the next section shows that sufficient inflation takes place in a large volume of the field space of $j$-inflation. 
It is in this context  of interest to investigate in more detail the behavior of the inflaton trajectories in different regions of the target space. 
The geometry and  the physics of the orbits depends of course on the potential surface above the target space. 
In particular the structure of the regions 
with sufficient inflation, or the existence of  attractors, is  dependent on the form of the potential, hence necessitates the 
consideration of inflaton trajectories on the potential surface.  The inflaton behavior can also be considered in the field space itself by 
projecting the inflaton paths down to the target space, 
where they can be compared  with the geodesics defined by the target space metric, in the present case the hyperbolic metric of the 
upper halfplane. The latter come in two types, straight vertical lines and semi-circles, and the natural question is how
 the inflaton trajectories compare to the geodesic paths. Naively one might expect the geodesic motion in the field space to be 
 irrelevant in slow-roll inflation because it is the gradient structure of the potential that is important and the dynamics of the inflaton 
 itself even along approximately geodesic paths will depend on the potential. However, it will become clear below that sufficient 
 inflation can be achieved with a 
 family of inflaton orbits that interpolate between the two types of hyperbolic geodesics.

\subsection{Hyperbolic geodesics and the effect of modular dynamics}

A natural question in the case of curved target spaces is how the physical geometry of the field space differs from the purely 
geometric dynamics, i.e. in what way the inflaton potentials affect the trajectories projected down to the target space as compared 
to  the  geodesics.  This is also of interest in the context of how the swampland conjectures might relate to the observational 
bounds on $r$ because the former are concerned in part with geodesic distances while the latter impact the 
distances along inflaton trajectories. The geodesic equations for the upper halfplane $\cH$ lead to the coupled system given by
  \bea
   \tau^{1''} - \frac{2}{\tau^2} \tau^{1'} \tau^{2'} &=& 0 \nn \\
   \tau^{2''} + \frac{1}{\tau^2} \left(\tau^{1'})^2 - (\tau^{2'})^2\right) &=& 0,
  \eea
  where the dimensionless variables $\tau^I$ are used and primes denote the derivative relative to the dimensionless time.
  These equations can be solved in terms of straight vertical lines $\tau^1=\rmconst$ as well as by semi-circles 
 that intersect the $\tau^2=0$ axis vertically, i.e. with centers along the real axis. 
 Parametric forms of these geodesics can be obtained in two different ways, using either a group theoretic 
 parametrization or a formulation  in terms of hyperbolic functions. 
  More precisely, the geodesic solutions can be given as a combination of a pair of maps of the point $\tau=i$ defined 
  by  elements $\b_t, g$ in the M\"obius group as 
  \beq
  \tau_g(t) ~=~ g \circ \b_t(i), ~~~~~\b_t ~:=~ \left(\matrix{e^{t/2} &0 \cr 0 &e^{-t/2} \cr}\right),
    ~~ g ~=~ \left(\matrix{ a &b \cr c &d\cr }\right) \in \rmSL(2,\mathR) 
 \lleq{group-geodesics}
 where the action is given as usual by the fractional transformation. The identity $g ={\rm id}$ leads to vertical geodesics 
 on the imaginary axis $\tau = e^t i$ and the shift matrices generate other vertical lines. This shows that 
 the upper halfplane $\cH$ is a  geodesically complete manifold. 
 The  group theoretic geodesics (\ref{group-geodesics}) 
 can be mapped into the hyperbolic function parametrization 
 that provide an alternative parametrization of the semi-circle geodesics.
 
   The selection of  $j$-inflation trajectories shown in figure 3  for several initial values illustrates that  
  for a wide range of initial values the   potential of $j$-inflation, and hence modular inflation in general, 
  has an important effect on the specifics of the dynamical solutions. Even with the constraint imposed on the 
  number of e-folds the $j$-inflation trajectories form a continuous family that turns from the vertical line type of 
  geodesics into the semi-circle type of geodesics. For initial values with large $\rmIm~\tau$ and $\rmRe~\tau$ 
  closer to the boundary of the central vertical band the trajectories approximate the vertical geodesics, 
  while for initial values closer to the saddle point  $\tau_s=i$ they eventually approximate the semi-circle geodesics. 
    In the process of this interpolation the family of trajectories inevitably contains paths that are far from being of 
  geodesic type.

If one imposes further phenomenological constraints such as the spectral index and the tensor ratio the $j$-inflation 
initial values are forced to start closer to the saddle points, as illustrated by a selection of different trajectories in figure 3.
These paths show that during the early phases of $j$-inflation even those trajectories that are compatible with the 
{\sc Planck} constraints for the spectral index and the tensor ratio can have interesting dynamical behavior that deviates 
from the geodesic. On the other hand  there do exist trajectories in $j$-inflation that approach the geodesic very quickly. 
This  comparison of course only concerns the geometric form of the paths involved. Physically, the potential surface is 
still important even for trajectories that are close to being geodesic because it is the potential that determines the 
time scales of the inflaton dynamics.
 \begin{center}
 \includegraphics[scale=0.1]{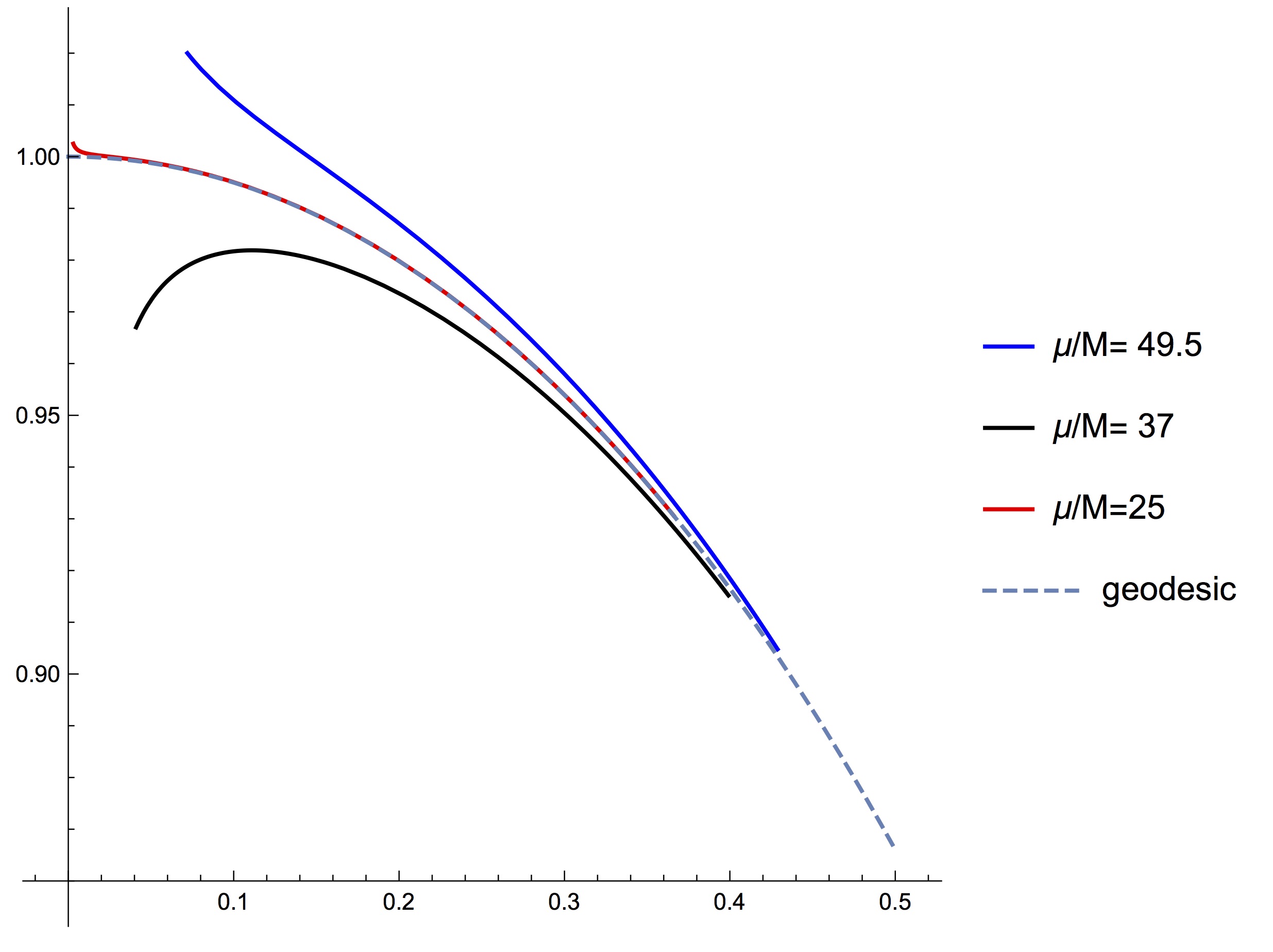}
 \end{center}
 \baselineskip=15pt
 \parskip=0pt
 \begin{quote}
 {\bf Figure 3.} ~An illustration of $j$-inflation trajectories for different energy scales $\mu$, and the geodesic that 
 connects the saddle point $\tau_s$ with the minimum of the $j$-inflation potential. 
  As before, these $j$-inflation trajectories are truncated at the end-of-inflation, but the geodesic is tracked to 
 the minimum of the potential.
 \end{quote}

\baselineskip=17.5pt
\parskip=0.1truein

Figure 3 also provides  another illustration of the attractor nature of the inflationary trajectories on a more local scale. 
If one continues to follow the dynamics along these 
orbits they all converge onto the geodesic path that connects the saddle point $\tau_s=i$ with 
the minimum of the $j$-inflation potential. This feature is enhanced for trajectories that originate close to the saddle point.

\vskip .2truein

\subsection{Attractor-like behavior}

In the context of the issue of fine-tuning of initial conditions it is natural to ask whether inflationary models show
 features that are reminiscent of attractor behavior for inflaton trajectories. 
 The mathematical concept of an attractor involves large time behavior \cite{m85}, 
hence strictly speaking it is not applicable in models in which inflation ends after a finite time, whatever the mechanism. 
It is nevertheless of interest to ask whether an inflationary potential has basins into which inflaton trajectories with quite 
different initial conditions tend to converge
and whether the trajectories that converge to these basins have similar physical parameters. 
For singlefield inflation there is a large amount of literature on this topic, and for  a special class
of models this is discussed briefly in the recent reviews \cite{b16, l17} and references therein.
In the framework of multifield inflation attractors have received less attention, but a recent discussion
can be found in \cite{fk18}. In light of these discussions it is of interest to consider trajectories in $j$-inflation.

The illustration of the potential in figure 1 shows that it increases and forms a wall 
for large imaginary components of the inflaton doublet. One can ask what the trajectories are that these initial conditions with 
sufficient inflation lead to and figure 4 shows a selection of $j$-inflation initial values and their associated trajectories.  
The paths in this graph all lead to inflation within the canonical range for the number of e-folds $N_* \in [50,70]$ 
adopted in the present paper, and they are tracked to the end of inflation.  
\begin{center}
\includegraphics[scale=0.35]{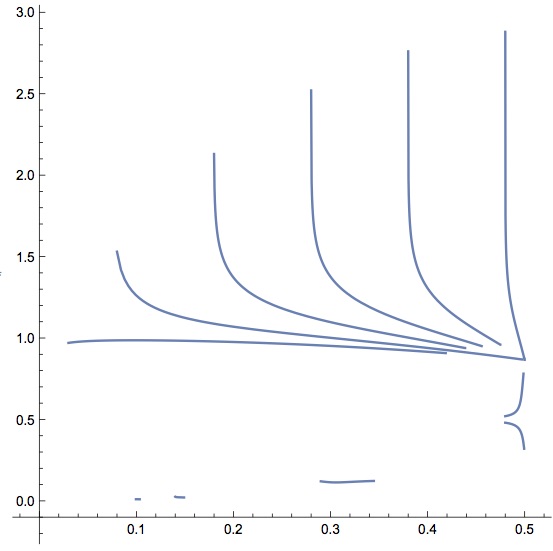}
\end{center}
\baselineskip=15pt
\begin{quote}
 {\bf Figure 4.} $j$-Inflaton trajectories with e-folds in the standard range $N_*\in [50,70]$. The hyperbolic metric 
    rescales the physical length of these paths drawn in the diagram.
   \end{quote}
 
\baselineskip=17.5pt

Included in figure 4 are also trajectories that start in the neighborhood of saddle points of the potential that are 
closer to the real axis. These trajectories appear to be much shorter in the coordinate space variables of this graph 
but because of the nontrivial 
hyperbolic metric of the field space the physical length of these paths is larger since they are closer to the 
horizontal axis. The factor $(1/\rmIm~\tau)$ in the length $ds$ leads to an ever stronger stretching 
as $\rmIm~\tau$ approaches zero.

The trajectories in figure 4 show that sufficient inflation can be obtained in $j$-inflation for a wide variety of different types of 
trajectories. Along some of these  orbits either of the two components $\tau^I$ of the inflaton can be approximately 
constant, which is an assumption that is sometimes made in the twofield inflation literature.
There are however also trajectories for which both of the components vary considerably, hence in $j$-inflation sufficient 
inflation can be obtained along trajectories that represent true twofield inflation.  

The paths shown in figure 4 also illustrate how the trajectories that originate close to the saddle point $\tau_s=i$ serve as 
an attractor basin into which a continuum of trajectories merge that start away from the saddle point. This behavior extends even 
to trajectories that start higher up the wall region of the potential. 
Only orbits that approach the linear type geodesics when projected down onto the field 
space approach the minimum in a more direct way. Their structure approximates vertical geodesics, which are 
briefly described above.

\vskip .3truein

\section{Sufficient inflation}

The question whether a given field theoretic model might be useful in an inflationary context is usually posed in a first 
iteration as a  constraint on the lower bound on the number of e-folds $N_*$. The original motivation for inflation leads 
to a rough estimate
 for $N_*$ and a number of analyses in different models involving an inflaton doublet have focused 
 on an analysis of the phase space that lead to sufficient inflation in this sense. A model that has received particular attention 
 over the years is hybrid inflation \cite{l93}, and  papers that report results of such scans include 
    \cite{l96etal, lv96, t97,  ml00, cr08, c09etal, c10}.  Inflation must end eventually and in principle such 
    scans of the phase space should impose also an upper bound for $N_*$.
 
The duration of inflation has received much attention in the literature, not only because it is the parameter that is most
 immediately relevant for the puzzles that initially played an important role in the introduction of the inflationary framework,
  but because it  has also  served as a focal point to provide more precision to the question of how probable inflation is,
  for example in the formulation of Gibbons and Turok  \cite{gt06}.
   It has turned out  to be difficult to constrain the number of e-folds precisely, in part because it depends 
 on the choices that have to be  made to account for the post-inflationary evolution.  
 The main uncertainty in determining a definite range is the reheating stage, which is poorly constrained observationally
   and which is complicated because of its preheating phase during the early stages 
 of this process. Despite much  work on issues related to preheating no clear estimates for the number of e-folds 
 and the reheat temperature have emerged.  Consequently, the range of e-folds that has been considered in the literature
 covers a wide range,  reflecting the lack of constraints on the energy scales that are allowed in this context, 
 reaching from a lower bound given by the nucleosynthesis scale of some $(1-10)\rmMeV$, to the GUT scale of 
 $10^{16}\rmGeV$.
Given a choice for the accepted range of the number of e-folds, i.e. a choice of the evolutionary scenario, 
 a key issue to address is whether the given model under discussion provides enough inflation in the adopted framework.
  In the absence of better constraints on the post-inflationary stages, 
 a canonical range for the number of e-folds $N_*$ between horizon crossing and the end of inflation 
 has emerged  that posits that  50 to 60 is a reasonable minimal value for $N_*$.  Such values were for example
  imposed as a lower bound  in the work on hybrid inflation.  In the present analysis of $j$-inflation more specific 
  bounds for the number of e-folds are implemented.

\vskip .2truein

\subsection{Initial conditions in $j$-inflation for sufficient inflation}

 In the present paper a conservative range for the number of e-folds is chosen as $N_*\in [50,70]$  to define
  viable realizations. Using this as a constraint on $N_*$, the existence of trajectories in $j$-inflation
 with enough inflation was established for $j$-inflation in \cite{rs14, rs16}. 
 (The generalization to higher level models  was considered in \cite{ls19}.)
 A systematic scan of the slow-roll phase space is computationally expensive  at high resolutions of the field 
 space, with most of the time spent on the  computation of the number of e-folds $N_*$.  
 A balance thus has to be found between the resolution of the lattice of the inflaton space as well as the time resolution 
   used  in the determination of  the end of inflation $t_e$.  By choosing the 
  inflaton resolution  $\delta \tau^I=\delta\phi^I/\mu$ low enough it is possible to scan the 
  central vertical band out to a range of $\rmIm~\tau$ where the e-folds fall outside 
  of the adopted range. A typical scan of the initial values leads to a structure that is reminiscent of a butterfly, 
  as shown in figure 5. 
  \begin{center}
 \includegraphics[scale=0.12]{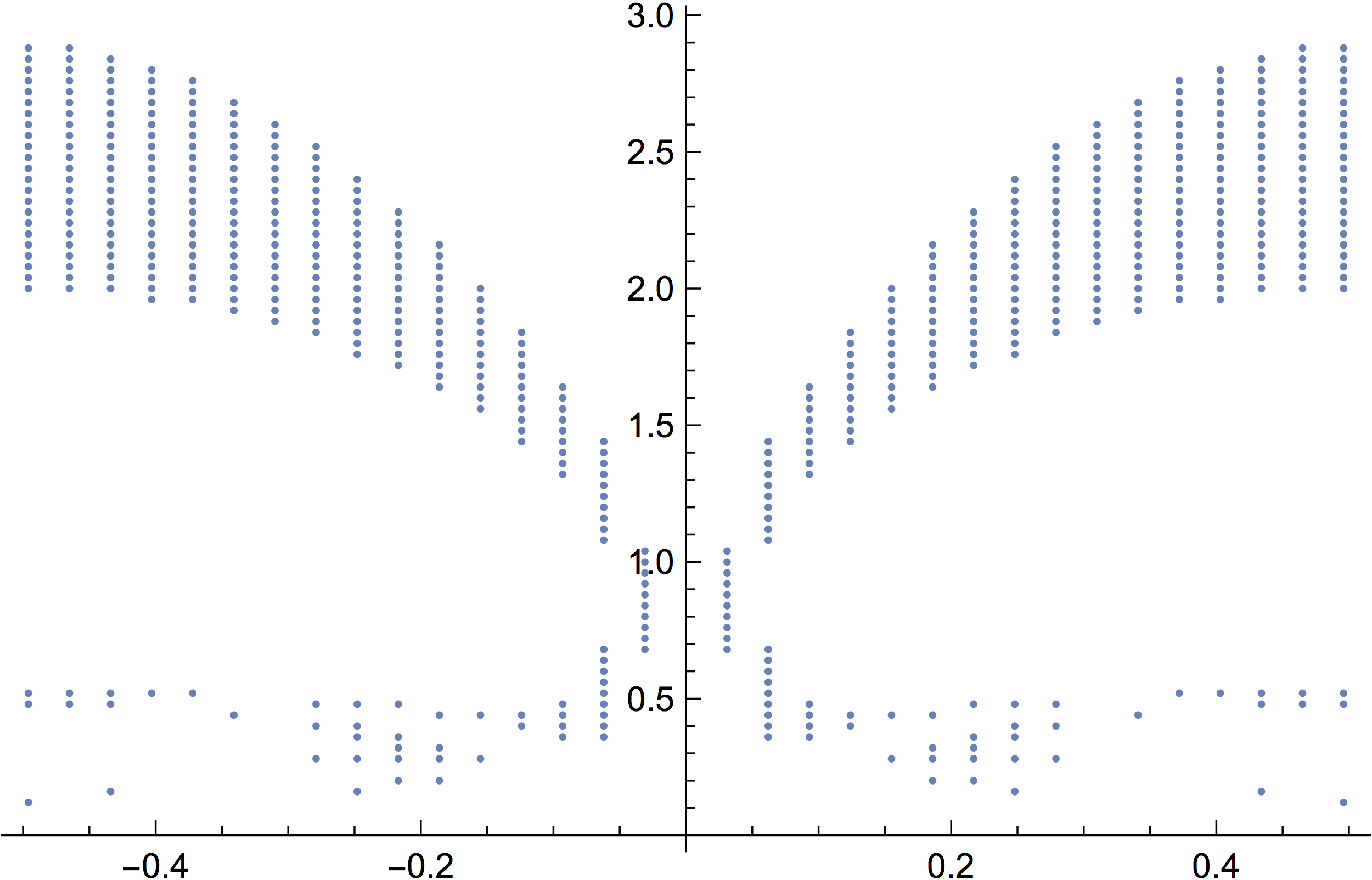}
 \end{center}
 \baselineskip=15pt
 \begin{quote}
  {\bf Figure 5.}~The butterfly graph determined by the constraints on the number of e-folds $N_*$. 
   \end{quote}

\baselineskip=17.5pt

\vskip .2truein

The feelers of the butterfly distribution reflect the intricate structure of the potential surface as one approaches the horizontal axis
$\phi^2=0$ that defines the boundary of the modular inflation field space. This structure  is indicated in the 3D plot of figure 1 
and can be seen more clearly in the global contour plot of figure 2.
A discussion of the behavior of the inflaton trajectories corresponding to these initial values has been given in the 
previous section. The structure of the 2D butterfly can be explained by resolving the graph by adding the number 
of e-folds $N_*$ as  a third dimension. Figure 6 shows that the  upper and lower boundaries of the butterfly structure 
are determined by the bounds on the number of e-folds. They are obtained because when moving away from  of the thorax 
 like structure that appears close to the ridge of the potential 
along the imaginary axis the e-folds produced by the potential are below the lower bound of the interval  $N_*\in [50,70]$ in the lower 
region of the target space and above the upper bound in the upper region of the butterfly wings.  Varying the bounds for the e-folds 
therefore varies the wing range  accordingly.
\begin{center}
 \includegraphics[scale=0.085]{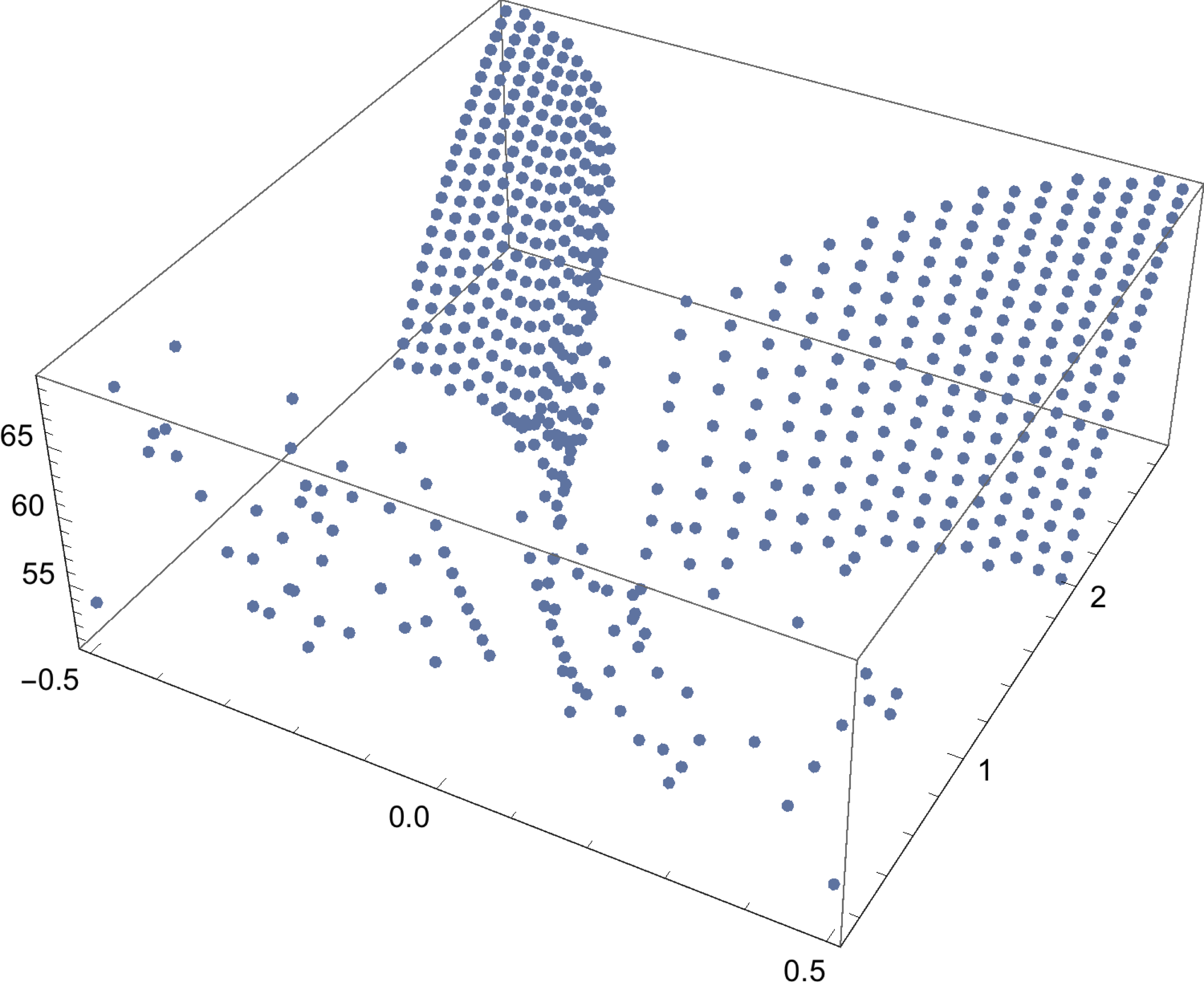}
  \end{center}
  \baselineskip=15pt
 \begin{quote}
 {\bf Figure 6.}~The 3D function of the e-fold number $N_*$ on the dimensionless target space $(\tau^1,\tau^2)$ for e-folds 
  in the range $[50,70]$. This graph shows that the upper and lower boundaries of the butterfly wings are determined by the 
       upper and lower limits adopted for the number of e-folds.
     \end{quote}
 
 \baselineskip=17.5pt
 
\vskip .3truein

\subsection{A natural measure on the space of initial conditions}

A question often raised in the context of inflation is how much of the space of initial conditions actually leads to inflation.
The answer to this question involves some kind of measure that needs to be chosen.
For general multifield inflation with a curved field space the target space comes equipped with a natural measure 
given by the volume measure $\om = \sqrt{\rmdet~G_{IJ}} \prod_K d\phi^K$. Even though in general the volume of a 
noncompact 
space will diverge in this measure, the situation is different in modular inflation, where the metric measure provides a 
finite volume for  the fundamental domain of the target space. There are two features of modular inflation that combine
 to allow the computation of a finite 
fraction of the initial conditions that lead to sufficient inflation. The first is that the modular symmetry constructs the full 
target space as an infinite number of copies of the irreducible fundamental domain $\cF$  of the upper halfplane $\cF\subset \cH$, 
given by
 \beq
  \cF ~=~ \cH/\rmSL(2,\mathZ),
 \eeq
 which extends infinitely high along the imaginary axis in the interval $[-1/2, 1/2)$. 
The second feature is that the metric is given by (\ref{hyperbolic-metric})  and while  the euclidean area of $\cF$ is infinite,
the hyperbolic metric  provides a finite area for the fundamental domain, given by
  \beq
  \rmvol(\cF) ~=~ \int_\cF \sqrt{\rmdet~G}~d\phi^1d\phi^2 ~=~  \frac{\pi}{3} \mu^2 ~=~ - \frac{2\pi}{3} \frac{1}{R_\rmfs}.
 \eeq
 It will become clear in the later sections that the CMB data constrains $\mu$ and hence the physical area of the fundamental 
 domain.

Given that the volume of the fundamental domain $\cF$ is finite,
 it is natural to consider the quotient of the butterfly volume and the total volume as a measure of the likelihood of $j$-inflation
 with a number of e-folds in the canonical interval.  This quotient can be interpreted as the probability  for $j$-inflation to 
 produce $N_*\in [50,70]$ e-folds. It  can also be interpreted as the posterior probability by assuming that the prior probability 
 distribution is constant, which allows to cancel it in Bayes' theorem. As a result modular inflation leads to the 
 probability
   \beq
   P(N_*\in \cI) ~=~ \frac{\rmvol(B_\cF)}{\rmvol(\cF)}.
  \eeq
  A rough estimate of the volume of $B_\cF$ shows that a large fraction of the total field space leads to inflation in the 
  range considered here.
  In the literature, the probability of inflation is often discussed by setting a lower cut-off for $N_*$, for example $N_*\geq 60$, 
  see e.g. the papers  \cite{l96etal, lv96, ml00, cr08, c09etal, ep13}, where the space of initial conditions with 
  $N_*\geq 60$ is analyzed for hybrid inflation.   The structure of $j$-inflation is such that if one lets $N_*$ grow 
  unbounded then the probability approaches one.
    
 \subsection{Finite measures of modular inflation and the swampland conjectures}
 
In the past the question of how likely inflation is has often been phrased in terms of measures that are not normalizable, 
 leading to regularization issues  \cite{ghs86, hp87, ct10, sw12}.  It is worthwhile to note here 
 that the conjectures that have been formulated in the context of the swampland conjectures place modular inflation 
 in a larger framework in which the existence of a geometric measure on the target space is guaranteed by fiat.
 
 The swampland conjectures have been formulated as an attempt to formulate criteria for effective field theories that can be 
 embedded in a theory of quantum gravity. For the most part it is assumed that this theory is string theory, and much of the 
 work aimed at providing evidence for the conjectures has been done within this framework. A review of recent work 
 can be found in \cite{p19} and further work on the impact of these conjectures on inflation includes \cite{wbh19}. 
 However, the formulation 
 of at least some of the criteria considered so far takes place within the effective field theory and does not directly refer to 
string theoretic features. Among the earliest constraints imposed were conditions extrapolated by Ooguri and Vafa \cite{ov06}
 from the behavior of string theory moduli.  These can be rephrased in terms of inflationary scalar
  fields, independent of any moduli interpretation. The focus of these early considerations was on the topological and 
  geometric structure of  target space of the 
  scalar fields, in general a multi-component field $\phi^I, I=1,...,n$.  The topological conjecture states that the 
  target space is non-compact in the sense that there exist trajectories of infinite length. The precise length of these trajectories 
  depends on the form of the distance measure \cite{rs18}, but this does not affect the basic picture of the 
   topological conjecture. Figure 3 illustrates the existence of such paths as one increases the imaginary range 
   of the inflaton doublet indefinitely.
  
  The geometric conjecture  posits that despite the non-compactness the target space has finite volume. 
  What is assumed implicitly in such statements is that the focus is on the fundamental domain. 
  In this context $j-$inflation, and more generally modular inflation, are examples of  theories that satisfy the 
  finite volume conjecture \cite{v05, dl05}, as discussed above.  

\vskip .3truein

\section{Spectral index function $n_{\cR\cR}$ on the field space}

One of the fundamental parameters that has been determined with a dramatic increase in precision by the 
post-COBE satellite experiments is  the adiabatic spectral index $n_{\cR\cR}$ of the curvature 
Lukash-Bardeen perturbation $\cR$. 
 The specific value and uncertainty of the spectral index depends on the details of the fits. 
 For the most part the fits that have been published by the collaborations are aimed at singlefield inflation 
 and the parameter space is correspondingly lower dimensional.  Fits that are aimed at twofield inflation  
 usually adopt special types of parametrizations for the 
 scalar power spectrum and often make assumptions concerning the correlations between the adiabatic 
 and the isocurvature perturbation. 
 Currently  no multifield specific fit of the final {\sc Planck} data release has been performed, but extended fits beyond 
 the minimal parameter count have been obtained in ref. \cite{dv19etal}.
 In the present paper bounds are used that are close to the {\sc Planck} values.

 In $j$-inflation the spectral index $n_{\cR\cR}(\tau^I)$ is given by the fairly complicated function in eq. (\ref{jinfl-params}) 
 in terms of the Eisenstein series. The analytic structure of this function is not immediately transparent but its geometry
 can be determined numerically and is shown in figure 7.
  The structure of this surface  explains some of the features of the numerical scans for viable initial conditions that 
  satisfy not only the constraint on the 
 number of e-folds, but are also consistent with the results from the CMB probes.
    \begin{center}
  \includegraphics[scale=0.07]{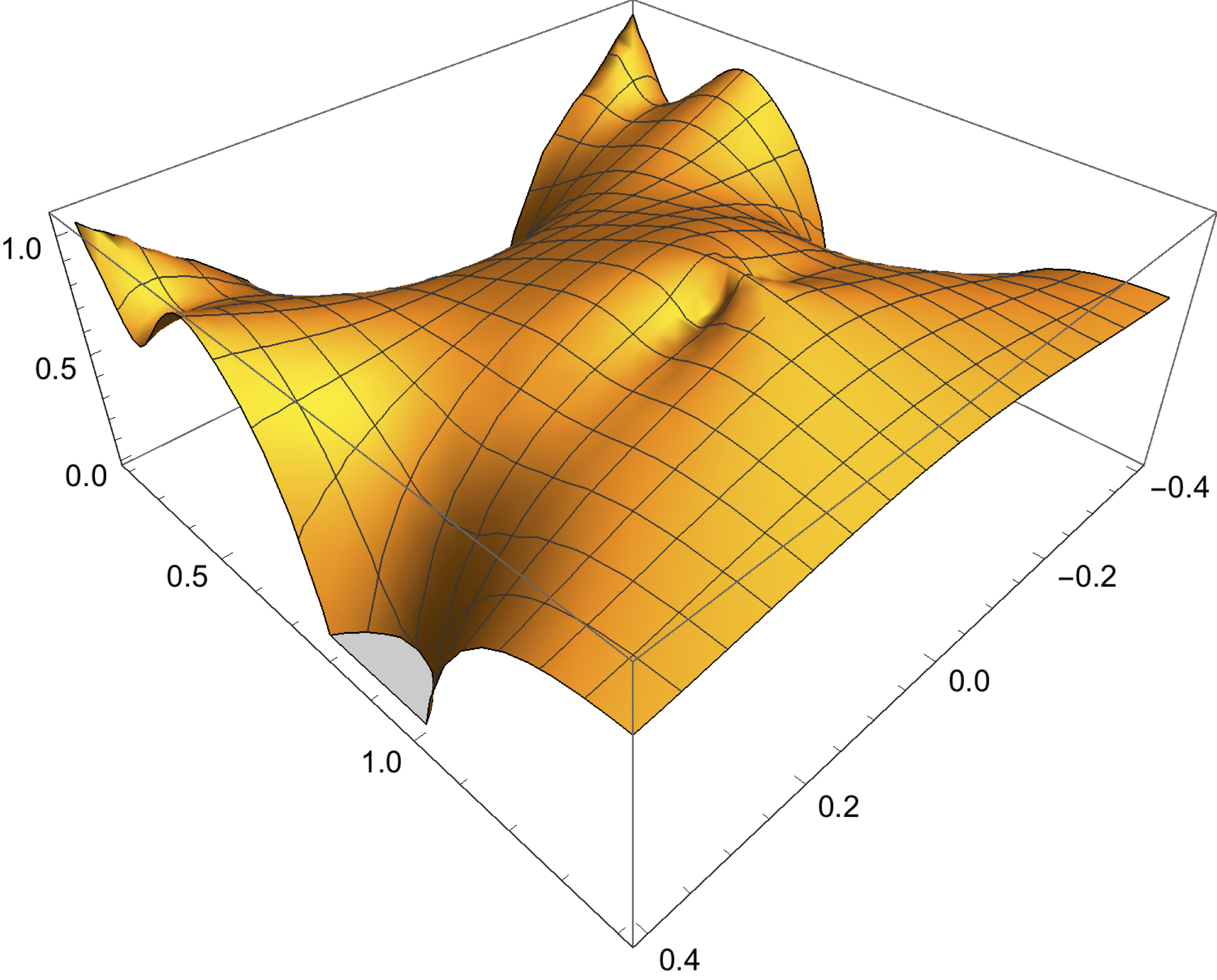}
  \end{center}
  \baselineskip=15pt
  \parskip=0pt
  \centerline{  {\bf Figure 7.} The spectral surface in a larger scale view.}
   
 \baselineskip=17.5pt
 \parskip=0.1truein
 
\subsection{$n_{\cR\cR}$ scans of the field space}

It follows from the $j-$inflation expressions for the spectral index and the tensor ratio in eq. (\ref{jinfl-params}) 
that the initial values in the
 neighborhood of the saddle points $\tau_s$ are good candidates that might lead to viable trajectories. 
 The region around $\tau_s=i$ was the focus of the phenomenological analysis of $j$-inflation of \cite{rs16}, 
 which established the existence of viable trajectories.
In the present section the focus is on the constraints imposed on the target space by the spectral 
index as the inflaton ranges over the target space for a fixed energy scale $\mu$. The effect of the variation of $\mu$ 
on the spectral index is considered later in this paper. The results of a global scan of the target space is shown in figure 8 
superimposed on the contour plot of the potential. This illustrates in some detail the 
canyon-like  structure of the field space close to the boundary. 
\begin{center}
\includegraphics[scale=0.6]{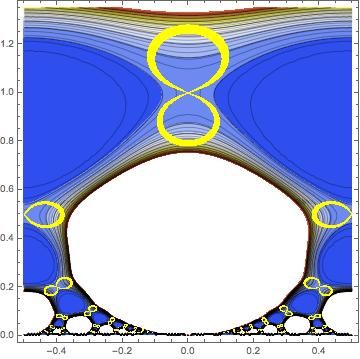}
\end{center}
\centerline{ {\bf Figure 8.} A scan for viable spectral indices on the field space $(\tau^1,\tau^2)$. }

\vskip .1truein

  This global map indicates a chain like iterative structure of saddle points which can be resolved only to a limited 
  degree in the region very close to the real axis. In order to obtain a more detailed view it is useful to construct a zoom 
  for smaller values  of the vertical component of the inflaton. 
  In figure 9 the focus is on the region to the right of the ridge, for which this higher resolution 
  graph shows a more detailed structure as one approaches the real axis,
   in particular a double sequence of arcs made of saddle point regions that 
   range from the right boundary of the central vertical band toward the origin of field space. These regions come in
   two different orientations, which are roughly diagonal and off-diagonal. Three different arcs are clearly visible and 
   a fourth one is indicated. The amount of detail visible in such runs depends on the size of the lattice used to scan 
   the field space, as well as the accuracy with which the Eisenstein series are computed. 
  \begin{center}
\includegraphics[scale=0.1]{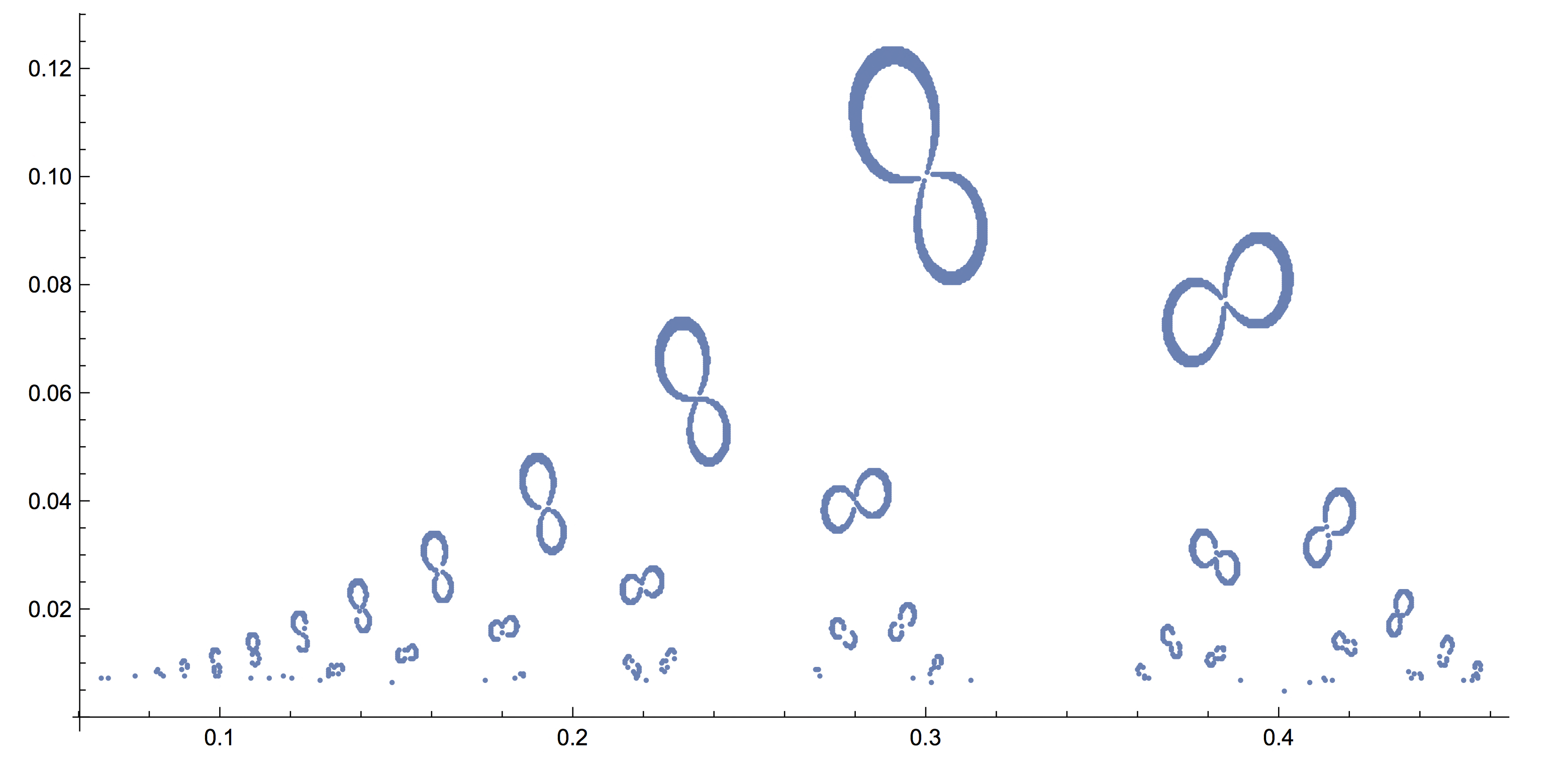}
\end{center}
 \centerline{{\bf Figure 9.} A zoom of the right part of the central region of the inflaton space $(\tau^1,\tau^2)$.}

\vskip .3truein

\section{Scaling behavior of the spectral index} 

A question that has been raised many times in the inflation literature is whether it is possible to derive constraints
for the amplitude of the gravitational contribution to the CMB background, given some plausible assumptions. 
The idea has been to abstract some general behavior from simple models that allows to deduce in particular 
a lower bound for the tensor-to-scalar ratio $r$
considered by the CMB collaborations \cite{planck18.6, planck18.10, a18etal}. One strategy that has been adopted
 in singlefield inflation is to assume some scaling behavior of the equation of state parameter and to consider the 
implications that result \cite{m13}.  
This translates immediately into an assumption for the scaling behavior of the slow-roll parameters 
and hence the 
spectral index and the tensor ratio. Given the CMB constraints on the spectral index this translates into 
estimates for $r$ \cite{m13, c14etal}.

In singlefield inflation such a scaling behavior is immediate in some simple models. In  the class of monomial inflation
 models \cite{t83},   the spectral index 
 \beq
  n_{\cR\cR} ~=~ 1 - 6\e_V + 2\eta_V
  \eeq
  simplifies because of the relation $\e_V = 2(p-1) \eta_V/p$, with $p$ the monomial exponent, 
  as well as the fact that the slow-roll parameter $\e_V$ can be 
  written in terms of the  number of e-folds as $\e_V \cong 1/N$.
 This leads to the scaling relation
  \beq
   n_{\cR\cR} ~\cong~ 1 - \frac{\a_p}{N}, 
  \lleq{ns-scaling}
  where $\a_p = (p+2)/p$. This relation also holds approximately for different classes of models, for example
  the Starobinsky model \cite{s80}.
  For about $N=60$ e-folds the resulting spectral index is consistent with the CMB constraints that 
  have been determined in recent years by the 
  WMAP and {\sc Planck} satellites. The question can be raised whether this scaling holds for some range of $N$
  away from the specific value $N=60$, often adopted as the canonical value.
  
  As noted above, such scaling relations were assumed to hold in attempts to put constraints on the undetermined, 
  but bounded, tensor-to-scalar 
  ratio $r$.   It was in particular observed by Mukhanov \cite{m13}  that the observational bounds obtained by the
   WMAP satellite for the spectral index \cite{wmap9},  in combination with a relation like (\ref{ns-scaling}),
 allow to determine lower bounds on the tensor-to-scalar ratio $r$ (see also the papers \cite{r13, c14etal}). 
 This raises the question whether such a relation might be valid more generally in the 
 framework of multifield inflation. In this case there are in general no analytical formulae for $N$ and hence there is no 
 easy access to analytical scaling relations. However, the analysis presented above of the behavior of $j-$inflation 
  allows to address this issue. 
  
  In order to test whether the above scaling relation for $n_{\cR\cR}$ holds more generally in multifield inflation
   it is useful to define the function
   \beq
    f_\a(n_{\cR\cR}, N) ~=~ n_{\cR\cR} - \left(1 -\frac{\a}{N}\right)
   \eeq
   for some  constant $\a$   and consider the correlation of this functions with the spectral index. If the scaling relation above holds the function 
   $f_\a$ vanishes.
  The analysis of the scaling relation in  $j$-inflation shows that the values of the function $f_\a$ for $\a$ of order one are small, 
  of the order of a few percent. This is illustrated by the graph in figure 10. 
While the function $f_\a$ is small, the behavior of the tensor ratio $r$ further above shows that the bounds on $r$ obtained in the 
framework of singlefield inflation are not valid in multifield inflation.

 \begin{center}
    \includegraphics[scale=0.08 ]{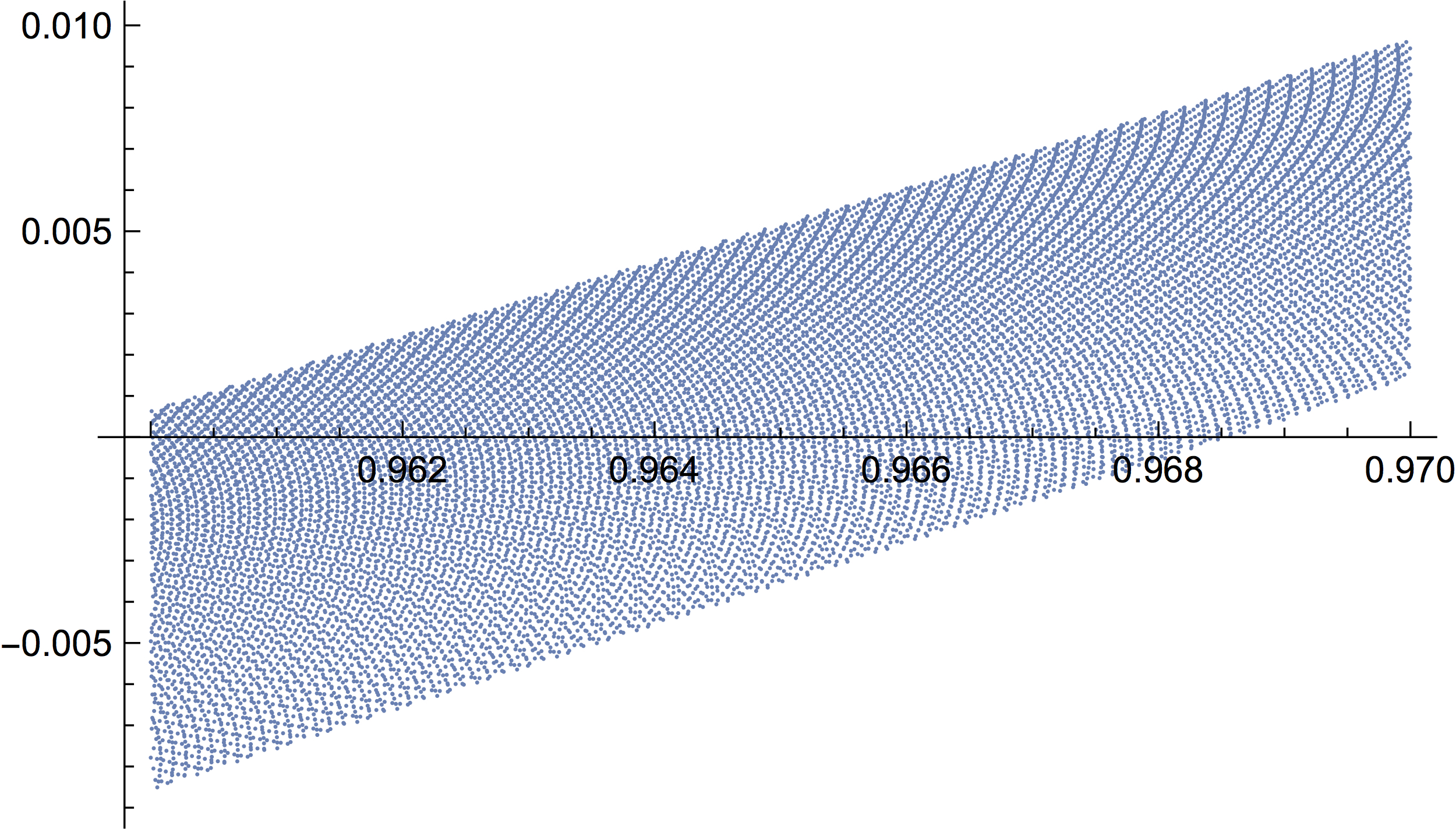}
 \end{center}  
\begin{quote}
{\bf Figure 10.}~~Typical graph for the function $f_\a$ in correlation to the spectral index $n_{\cR\cR}$.
\end{quote}

\vskip .3truein

\section{The tensor-to-scalar ratio function $r$ on the field space }

One of the fundamental phenomenological parameters that constrain inflation is the power amplitude of primordial  gravitational waves. 
This is conventionally quantified by the CMB collaborations in terms of the tensor-to-scalar ratio $r$ defined in 
eq. (\ref{params-def}).  For $j$-inflation this parameter becomes a function of the energy scale $\mu$ that determines 
the curvature of the target space and the inflaton field $\phi^I$ as in eq. (\ref{jinfl-params}). This observable $r$ has not yet  been 
determined, but the bounds of the amplitude of the gravitational perturbations have dramatically improved since COBE and further 
 experiments to constrain $r$ are currently under construction. The main point of these efforts of course is to detect primordial 
 gravitational waves, and in the process determine some of the characteristic features of inflation.  However, while the scale
 of inflation is an important characteristic, it is not the only information tied to $r$ and even in 
 the absence of a detection the  results of future experiments are important   because  they provide essential
 constraints for inflationary model building. While an improvement of an order of 
  magnitude in the recent past has not provided insight into the scale of inflation \cite{l96}, the experimental achievements 
  have been strong enough to exclude an infinite number of models, among them the class of 
  monomial inflation potentials.  
 
 \subsection{Large field inflation and small tensor-to-scalar ratios}
 
 One of the issues in which the bounds on $r$ have been instrumental is the question whether the experimental 
 target values predicted to be reached in the near future will be able to exclude certain types of inflationary models.
  As noted above,  this has already been achieved by the exclusion of certain sequences of models.
 A more far reaching discussion that has been conducted in a  large number of papers, following \cite{l96}, is concerned with 
 the question whether there are bounds that can exclude large field inflation altogether. 
  In the past strong bounds have been suggested based on restricting consideration to special classes of inflationary models,
  and some of these bounds have been used in proposals for new observatories. 
   It was in particular  noted early on that a non-detection of $r$ at the level of $10^{-2}$ \cite{b08etal} or 
  $10^{-3}$ \cite{cl13}   would rule out large field inflation, see also \cite{ec06}. This value was subsequently reduced 
   somewhat,  depending on  specific assumptions made in the analysis. 
   Slightly lower constraints than the milli-scale mentioned above 
   were obtained in \cite{g14etal, h15} in  the context of singlefield inflation. For a special class of models 
   this bound was pushed down further in \cite{l16}, where it was argued  that large field inflation is only possible if 
   this ratio is larger than $r\geq 2\times 10^{-5}$.   In light of these discussions it is of interest 
   to analyze the tensor-to-scalar ratio  $r$ in the more general context of multifield inflation with curved target 
   spaces. 
      
   The current experimental bound for $r$ is several magnitudes above the most stringent values values just quoted. 
   An often cited value is the result reported by the {\sc Planck} experiment, 
   which is reported to be given by $r \leq 0.06$ \cite{planck18.6, planck18.10}.
 Such bounds depend on the type of fit adopted in the experimental  analysis and the above value was obtained 
 in a fit by assuming a minimal number of parameters, which is not appropriate for multifield inflation. 
 Extensions beyond the standard six parameter fits have been considered, but at present no {\sc Planck} based 
 full analysis appropriate  for twofield inflation is available in the literature. In the following the singlefield bound 
 will therefore be adopted as the constraint. Ongoing experiment such as CLASS \cite{class18, class21} 
 aim to reduce the {\sc Planck} bound to the level of $10^{-2}$, while upcoming and future  ground based 
 experiments  such as the Simons Observatory, the BICEP Array,  and the CMB-S4 project 
 will be able to reduce this by more than an order of  magnitude 
 \cite{h18etal, simons18etal, simons19etal, cmbs4-19a-etal, cmbs4-19b-etal, cmbs4-20etal}. 
 The satellite  experiment Light BIRD \cite{lightbird19} is designed to reach down to 
    $r=2\cdot 10^{-3}$ at 95\% C.L.,
  while the target value of  the proposed satellite experiment  PICO \cite{h-pico19etal, a-pico19etal} reaches further
  down by almost another  order of magnitude to  $r = 5\cdot 10^{-4}$ at a $5\si$ confidence level, 
  with $\si(r) = 1\cdot 10^{-4}$ at $r=0$.
   As noted already, even a non-detection of  the tensor modes at this level would have significant effects on the current 
   inflationary model  landscape  and it would  provide a better perspective on different ad hoc selection rules 
   that have been used to provide a lower bound on $r$ \cite{b05etal}. 

The goal in the present section is to analyze the  tensor-to-scalar ratio of $j$-inflation in light of the experimental and theoretical
 bounds discussed above. The systematic results are of course limited by the resolution of the scans, but even so it becomes 
   clear from the analysis below that the parameter space of $j$-inflation is constrained by the {\sc Planck} bounds on $r$, 
   hence $j$-inflation presents a target for upcoming  and planned experiments. It will also become clear that there 
   are regions in parameter space in which the lower bound of $r$  can be pushed below the experimental floor of all upcoming 
   experiments. The $r$-range of $j$-inflation in particular reaches below the boundary that has been identified in the 
   literature as the dividing line for large field inflation in the singlefield  context.
 The fact that $j$-inflation can reach below these proposed large field boundaries  alleviates the concerns 
 expressed in a number of recent discussions \cite{simons19etal, cmbs4-19b-etal, h-pico19etal, a-pico19etal} 
 that the measurement of an experimental bound for the tensor ratio below
 $r\cong 10^{-3}$ would force a significant change in our understanding of the primordial universe. 
 The example of $j$-inflation shows that the framework of twofield inflation is able to reach below these thresholds. 
 
 \vskip .1truein
 
 \subsection{Global $r$-scan} 
 
As in all scans of the target space the specific results obtained for the tensor ratio depend on the size of the fundamental 
lattice cell as well as the parameter $\mu$ that is not fixed by the CMB amplitude in the formulation considered here. 
What does not change is the global structure of the results. As in the case of the spectral index, it is useful to consider the 
 shape of the analytical form of the  tensor ratio $r(\tau^I)$ of $j$-inflation as a function on the target space via 
 eq. (\ref{jinfl-params}).  This is shown in figure 11.
 \begin{center}
  \includegraphics[scale=0.13]{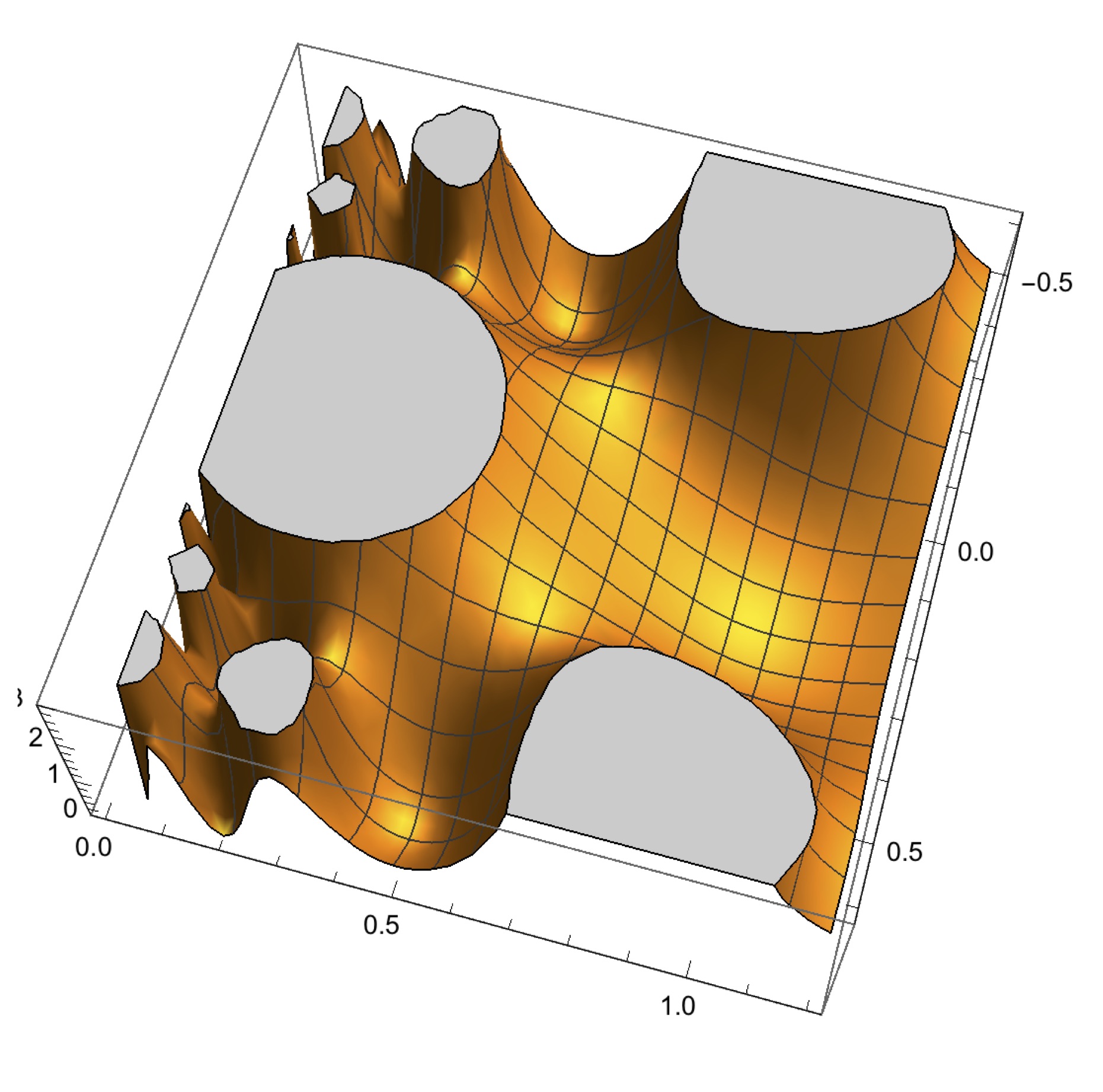}
 \end{center}
 \centerline{{\bf Figure 11.} ~Graph of the tensor-to-scalar ratio $r(\tau^I)$ of eq. (\ref{jinfl-params}).}

\vskip .2truein

In figure 12 the results are given for a global scan of $r$along the central vertical band.  
 The resulting pattern of this band is repeated in the upper halfplane  to the left and right by the 
 shift symmetry $\tau ~\ra~  \tau+1$ that is part of the modular group $\rmSL(2,\mathZ)$ that leaves the potential invariant. 
 It is this  invariance that is part of the motivation of modular, and more generally automorphic inflation \cite{rs14, rs15}. 
 The structure of the central band changes in dependence of the distance to the real axis. The areas closer to the boundary 
 of the target space given by $\rmIm~\tau=0$ are distorted in the plot of figure 12  because it is 
 drawn by using the 
 euclidean metric. The hyperbolic metric that describes the actual geometry of the field space increasingly
 stretches these domains as the horizontal boundary is approached 
 because of the factor $1/(\rmIm ~\tau)^2$ that enters $ds^2$.
  This measure leads to finite volumes of the target space in which the {\sc Planck} bound on the tensor ratio is 
  satisfied.
\begin{center}
\includegraphics[scale=0.14]{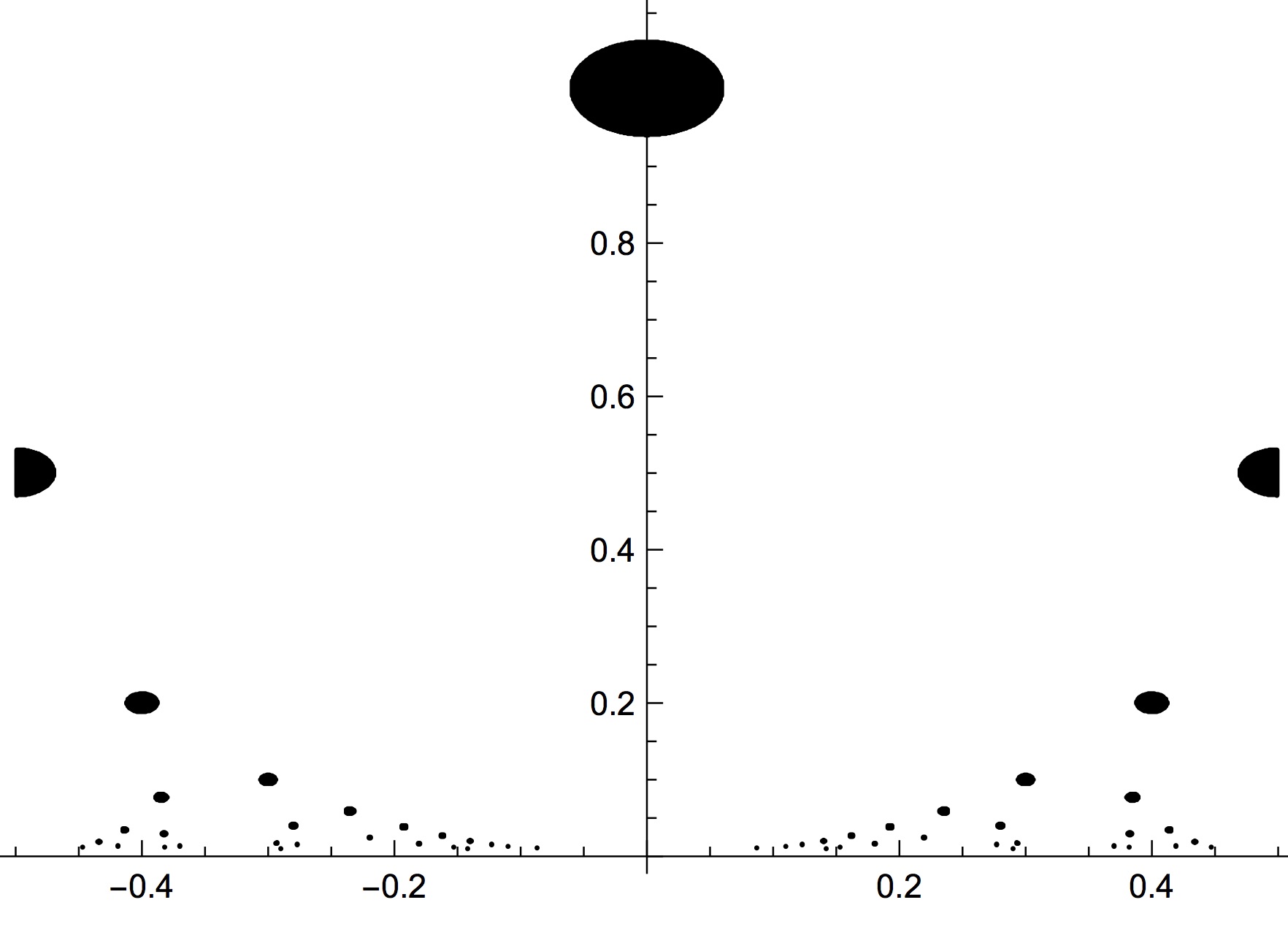}
\end{center}
\baselineskip=15pt
 \begin{quote}
 {\bf Figure 12.} ~ Global scan for viable initial values along the central vertical band consistent with the tensor ratio constraint.  
Close to the horizontal axis the effect of the hyperbolic metric leads to a stretching of the euclidean distances of
this plot.
 \end{quote}

\baselineskip=17.5pt

 \vskip .2truein
 
 \subsection{Effect of the field space curvature $R_\rmfs$ on the tensor ratio $r$}

Recall from the discussion above that once the energy scale $\mu$ is chosen, the overall scale $\L$ of the potential
follows from the CMB amplitude, making $\mu$ the only parameter that can be varied. This means that the only free 
parameter in the model is the curvature scalar of the target space $R_\rmfs=-2/\mu^2$ determined by the metric, 
 eq. (\ref{field-space-curvature}). The tensor ratio in eq. (\ref{jinfl-params}) shows that $r$ scales like $1/\mu^2$,
 hence it scales with the scalar curvature $R_\rmfs$ of the field space
 \beq
  r ~=~ -64 \pi^2 M_\rmPl^2~R_\rmfs~(\rmIm~\tau)^2 \left|\frac{E_6}{E_4}\right|^2.
 \eeq
 The second ingredient that varies is the inflaton value 
 $\phi_*^I$ at horizon crossing and these two parameters determine the range of the $r$. The precise boundary values
 of $\mu$, or $R_\rmfs$, and $\phi_*^I$ depend on the range of the number of 
 e-folds, and on the lattice resolution of the scan.

To be concrete, for a fixed value of the target space curvature $R_\rmfs$ via a choice of $\mu$ the initial values $\tau^I_*$ 
that satisfy the constraints adopted for $r, n_{\cR\cR}$ and $N_*$ sweep out  a finite band in the  $(n_{\cR\cR}, r)$-plane.
As the target space curvature is varied this band moves up and down along the axis defined by  the tensor ratio $r$. 
It is a priori not transparent how the imposed constraints  combine to determine the specific boundaries and the band in 
the $(n_{\cR\cR}, r)$-plane. This is illuminated by considering a three-dimensional plot which adds the number of 
e-folds $N_*$ to this plane.  Figure 13 shows that doing so explains the boundaries 
as determined by the bounds adopted for $N_*$. This lifting of the planar plot 
 shows that the lower bound on $r$ arises from the upper bound of the number of e-folds.  If the upper limit $N_*$ were 
 to  be increased  the lower bound on $r$ would decrease for fixed $\mu$.   The degeneracy of the sheet in figure 13 
 along the direction of the spectral index shows that in $j$-inflation 
 a scaling relation exists between the tensor ratio $r_*$ and the number of e-folds $N_*$.
Parametrizing the tensor ratio as $r = \a N^{-\b}$ one obtains for $\b=1$ the monomial class, 
for $\b=2$ the Starobinsky type range of $r$, 
and for higher $\b$ the range covered by $j$-inflation, with increasing suppression obtained for
 larger target space curvature $R_\rmfs$.
\begin{center}
\includegraphics[scale=0.15]{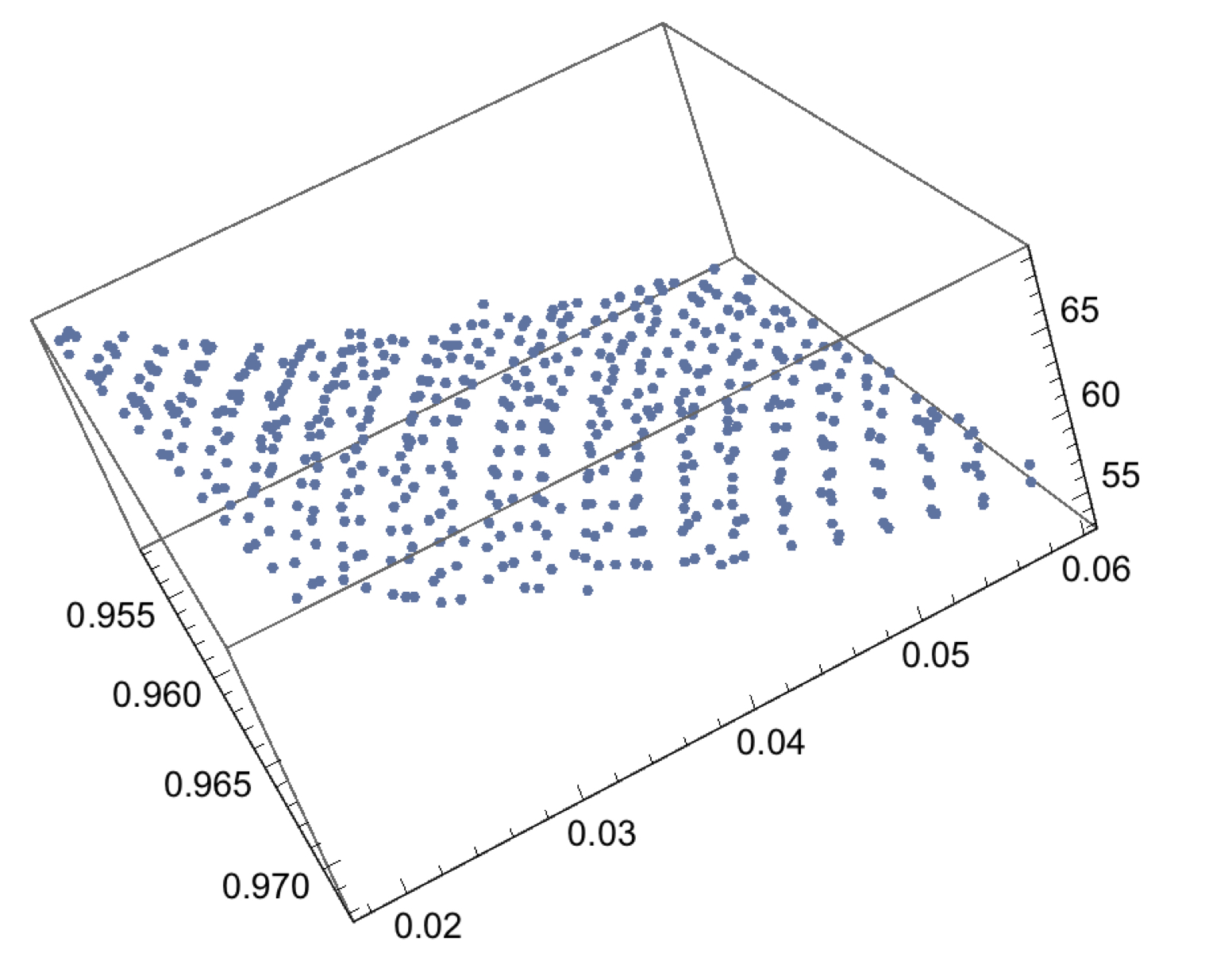}
\end{center}
\baselineskip=15pt
\begin{quote}
 {\bf Figure 13.}~ The $N_*$ dependence in the $(n_{\cR\cR},r,N)$ space at fixed $\mu$:  
 as the number of e-folds grows the tensor ratio $r$ decreases. 
 \end{quote}
 
 \baselineskip=17.5pt
 
 \vskip .2truein

The issue of the upper limit for $N_*$ has a long history and the specific bound $N_*\leq 70$, while often adopted, 
is not sharp. Discussions of the allowed range of the number of e-folds during inflation can be found in 
\cite{dh03, ll03} and their implications for trans-Planckian physics has also been extensively discussed. 
A review of these issues can be found in \cite{b16}.

As noted above, the band swept out in the plane spanned by $n_{\cR\cR}$ and $r$ in dependence of the CMB constraint on 
the spectral index and the adopted range for the number of e-folds depend on the energy scale $\mu$ that determines the 
field space curvature $R_\rmfs$. As the bounds obtained by the {\sc Planck} probe will be improved 
by the next generation of experiments currently under construction the bound on the curvature scalar will become 
stronger. This raises the question how much the tensor ratio is affected by a variation of 
the target space curvature that remains consistent with the {\sc Planck} bounds on the spectral index and $N_*$.  
In figure 14 the results for $R_\rmfs = 1/\mu^2$ with $\mu=27M_\rmPl$ and $\mu=35M_\rmPl$ are shown. 
This graph illustrates that the $r$-bands determined by 
the target space curvature  lie for both energy scales $\mu$ within the confidence region of the {\sc Planck} results.
As the curvature $R_\rmfs$ increases the band moves lower into the region targeted by the 
Simons observatory \cite{simons19etal}.  This shows 
that the range of the tensor ratio obtained in $j-$inflation for different $\mu$ contains and extends the region of the 
$r$-value obtained within the class of monomial inflation, as well as that of the Starobinsky model \cite{s80}. 
The latter in particular is a prominent model that has motivated some experimental proposals in the recent past. 
\begin{center}
 \includegraphics[scale=0.11]{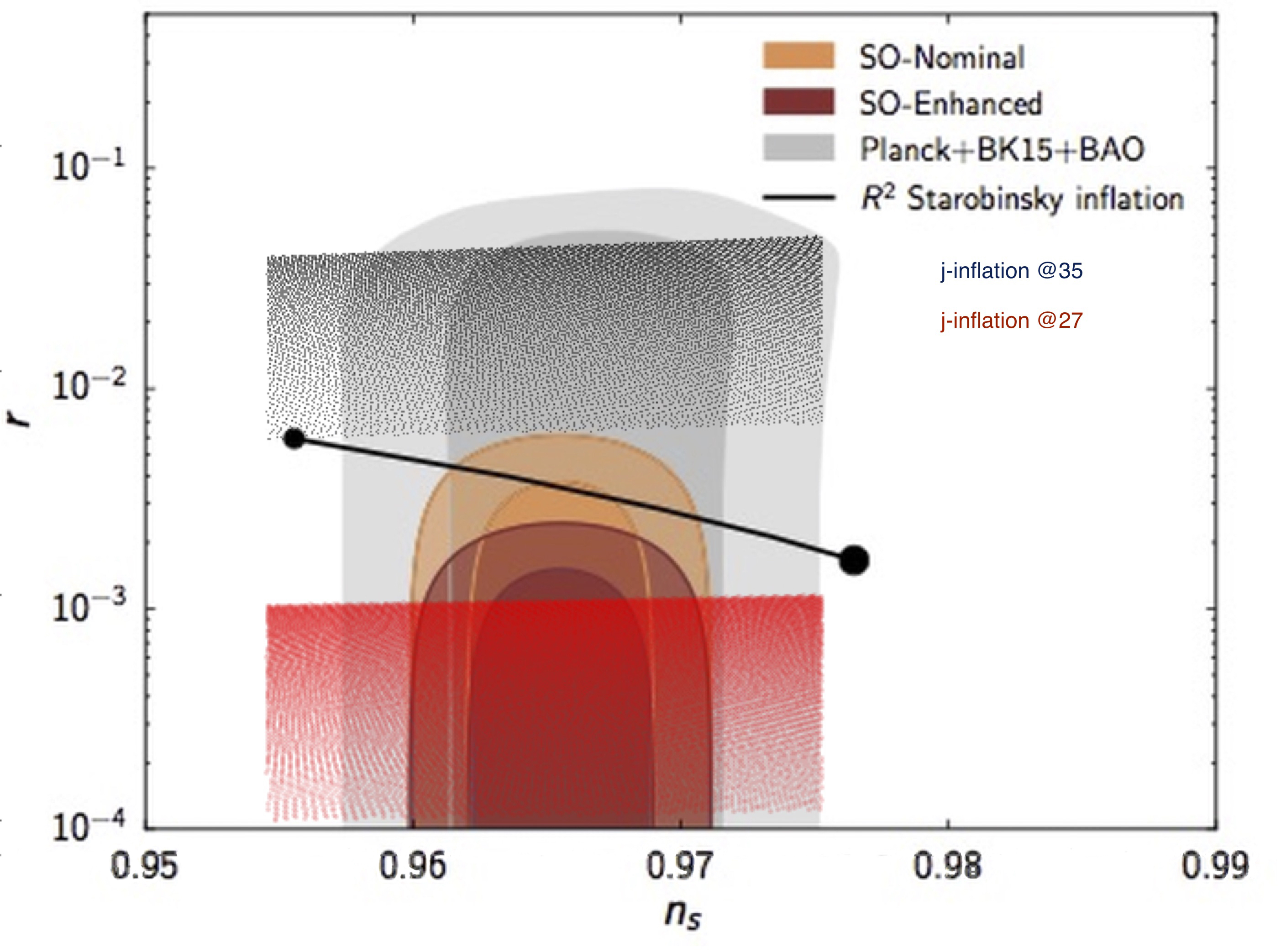}
 \end{center}
\baselineskip=15pt
\begin{quote}
{\bf Figure 14.} The results obtained for two scans of $j$-inflation  at energy scales $\mu/M_\rmPl=27,35$, 
with associated curvature $R_\rmfs$  of the  target space, projected onto the Simons Observatory confidence 
 regions \cite{simons19etal}.  
 The results for the larger $\mu$-value (black, upper region of the data points)  intersect with the 
 {\sc Planck} range, but not with the target range of the Simons telescope. For the smaller $\mu$-value
 (red, lower region) the  distribution intersects with the region described by the Simons Array collaboration. 
 \end{quote}
 
\baselineskip=17.5pt

The analysis of this section shows that  $j$-inflation provides an interesting target for 
experiments that aim to discover the gravitational contribution of the CMB signal, or at least to significantly
 improve the constraints on the tensor ratio. This includes experiments that are scheduled to come online in the near future, 
such as the Simons observatory, CMB-S4,  as well as the satellite experiment LightBird. 
The results from these observations will have an impact on the size of the field space curvature 
of the hyperbolic field space of modular inflation.

 \vskip 0.3truein
 
 \section{Effect of the $\mu$-variations}
 
Modular inflation as described in section three is parametrized by the  energy scale $\L$ and the inflaton scale $\mu$ needed 
to make the inflaton dimensionless. The scale $\L$ of the potential is as usual determined by the amplitude of the CMB 
power spectrum measured by the satellite experiments.  This leaves the scale $\mu$ as an a priori undetermined 
parameter of the theory, which is related to the field space scalar curvature via eq. (\ref{field-space-curvature}). 
This parameter is constrained however by the CMB results for the observables, such as the spectral index and the
 bound on the tensor ratio. Any scan of the target space is of course limited by the adopted lattice resolution and 
 other run parameters that are constrained by CPU resources. The scans discussed were 
mostly performed with a fixed $\mu$ and the question arises how these change when $\mu$ is varied. This is illustrated 
for the spectral constraint $n_{\cR\cR}$ in figure 15, which represents a
compilation of  initial inflaton values that satisfy the {\sc Planck} constraint for the range of $\mu\in [25,50]M_\rmPl$. 
The modular symmetry of the $j$-inflation model implies that the same structure is repeated along 
the modular sequences described in sections four and five, repeating this pattern indefinitely.

\begin{center}
 \includegraphics[scale=0.09]{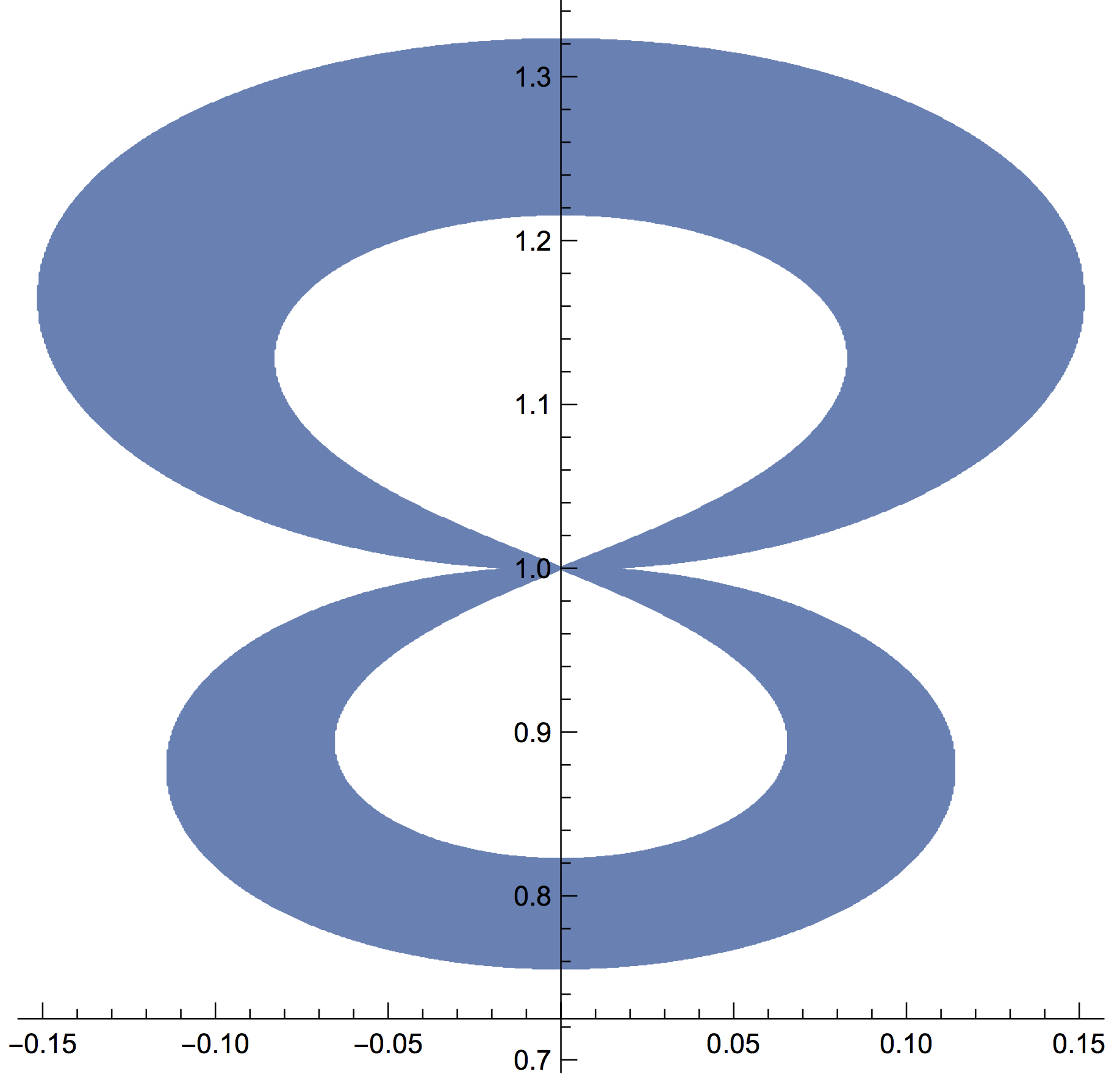}
 \end{center}
 \parskip=0pt
 \baselineskip=15pt
 \begin{quote}
 {\bf Figure 15.} A compilation of  results for the spectral scalar index for $\mu/M_\rmPl$-variations between 25 and 50
 in the region around the saddle point $\tau_s=i$.
 \end{quote}

\parskip=0.1truein
\baselineskip=17.5pt

Imposing not only the spectral index, but also the tensor-ratio bound of {\sc Planck} and the constraint 
on the number of e-folds $N_*$  selects from the figure eight of the previous graph a subset of initial values 
that for the same range of $\mu$-values and the same run-parameters
produces a region in the target space of the form of a nutcracker, shown in figure 16.
 \begin{center}
 \includegraphics[scale=0.13]{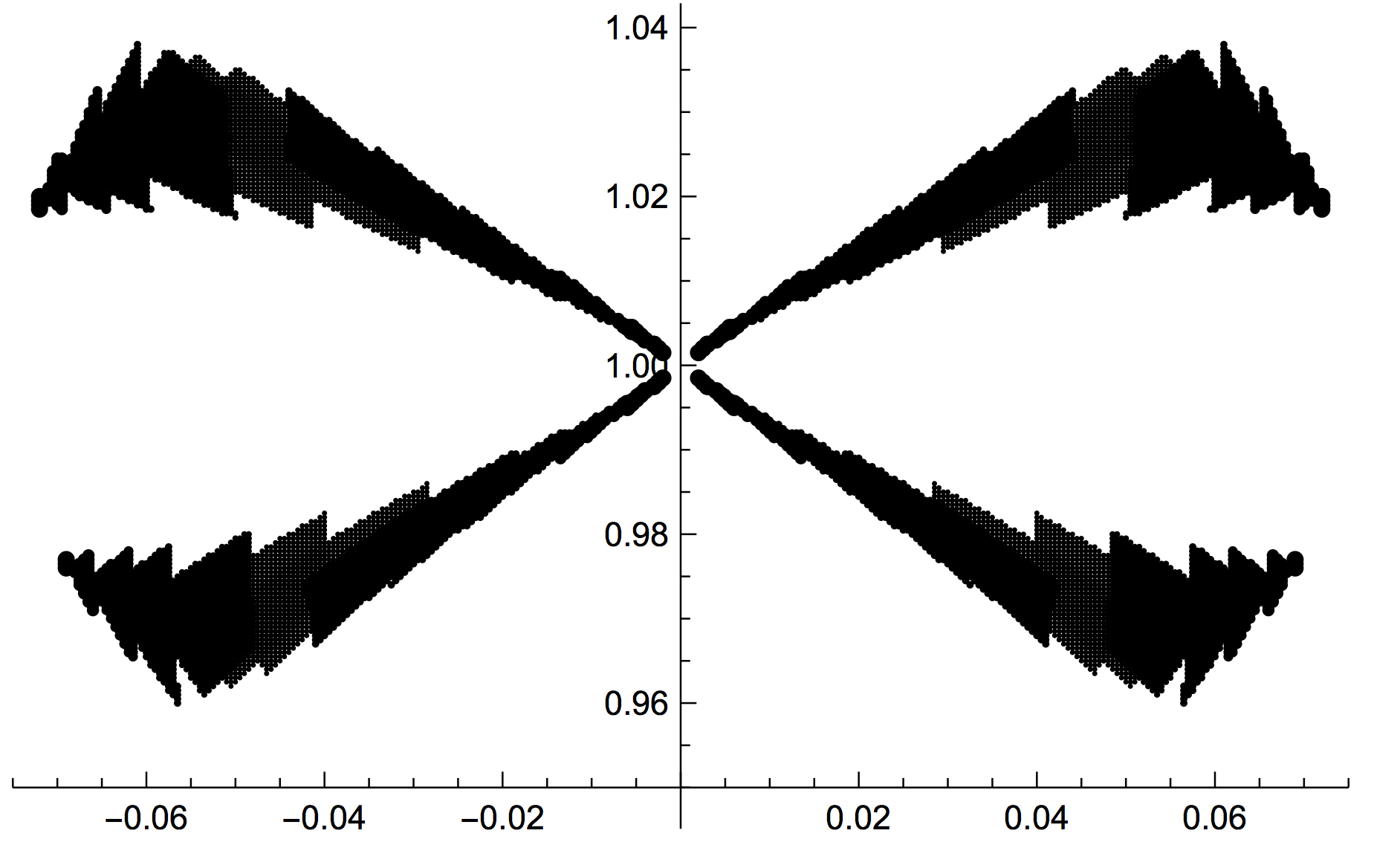}
 \end{center}
 \baselineskip=15pt
 \begin{quote}
  {\bf Figure 16.}~ The nutcracker of the constraints $n_{\cR\cR}\cap r\cap N_*$ for the runs in the previous graph,
  i.e. $\mu/M_\rmPl \in [25,~50]$.  
  \end{quote}

\baselineskip=17.5pt

A natural question that can next  be raised is how the tensor to scalar ratio $r$ behaves as the energy scale is varied. 
In the context of the $(n_{\cR\cR}, r)$ plane usually considered by the CMB collaborations this has been discussed
in section ten and illustrated in figure 14.  In figure 17 this is addressed 
by considering the 3D-distribution of the parameters $(n_{\cR\cR}, r, N_*)$ for three different values of $\mu$. 
This plot allows to extrapolate 
the structure for the intermediate values and also how the pattern extends beyond the values shown. The different 
bands in dependence of $\mu$ show that the scaling relation mentioned in the previous section between $r_*$ and $N_*$ 
varies with $\mu$. Figure 17 shows that the scaling relation mentioned already  depends on the energy parameter $\mu$. 
\begin{center}
 \includegraphics[scale=0.3]{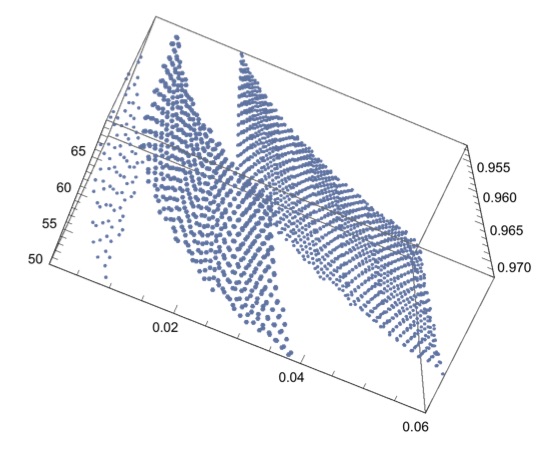}
 \end{center}
 \baselineskip=15pt
 \begin{quote}
 {\bf Figure 17.} Illustration of the $\mu$-dependence of the $(n_{\cR\cR},r,N_*)$ distribution for 
  $\mu ~\in~ \{30,35,40\}$.  The sheets move from right to left with decreasing $\mu$.
 \end{quote}

\baselineskip=17.5pt
\parskip=0.1truein

\vskip .3truein

\section{Conclusions} 

 The present paper has conducted an extensive systematic analysis of the inflaton behavior in modular inflation 
 with the potential given by the $j$-function. 
 In the process it has become apparent that the field space has an interesting structure that originates 
 from the underlying symmetry, which allows to perform a quite complete scan of the total target space, 
 due in  part to the shift-symmetry that is part of the modular invariance. The shift symmetry in particular 
 makes it possible to effectively scan completely the part of the target space constrained by the {\sc Planck} data
 because it leads to a decomposition of the target space into an infinite number of vertical bands that all
 have the same field theoretic behavior. 
  Within each of these bands the tessellation induced by the modular invariance in turn leads to an iterative structure 
   that maps large field regions to 
  regions where the initial field components take values that are arbitrarily small. 
  This shows that modular invariant inflation is a framework in which the distinction between large field and 
small field inflation into separate classes no longer holds. As a result the notion of 
large-field and small-field inflation is not a model characteristic of general inflationary models. 
While this has been shown here in the context of a particular model, the modularity admitted
by this type of geometry generalizes to other models in this class of theories. 
In the more general context of automorphic inflation with an arbitrary number of inflaton components 
a similar symmetry structure arises where the modular groups are replaced by discrete groups 
$G(\mathZ)$ for reductive groups $G(\mathR)$ \cite{rs14, rs15}.  As a result the classification of inflation should in 
general proceed along lines that do not reference the dichotomy of large field vs. small field inflation. 
A further implication of the tessellation of the field space in combination with the structure of the $j$-inflation 
potential is that in the 
geometric measure provided by the target space metric the fraction of the volume of the field space that leads 
to sufficient inflation is finite.

  A characteristic  feature of the potential surface of $j$-inflation is a wall-like structure of the potential along the vertical direction
  of the inflaton doublet.  The motion of the inflaton along this wall is of geodesic type, and is restricted by the 
  phenomenological constraints. Inflation trajectories that are compatible with the {\sc Planck} data exist in neighborhoods 
  around the critical points of the potential, leading to inflaton orbits that in general are different from hyperbolic 
  geodesics, but can approximate them as the initial values get closer to these critical points.
  As a result, while the presence of the hyperbolic metric in modular inflation has important implications for the structure 
  of the theory, the dynamics determined by the potential is crucial for the behavior of the inflaton.
   In particular the attractor structure of the model is determined by 
the potential, as expected. Nevertheless, there are regions in the target space where the geometry of the inflaton trajectories 
follows closely that of the different types of geodesics that characterize the upper halfplane hyperbolic target space. 
In the context of the dynamical structure the attractors in the potential surface project down 
to trajectories that are close to the two different types of geodesics that are encountered in hyperbolic geometry.

The phenomenological part of the analysis first shows that the region of the target space with sufficient inflation 
in the canonical range has the shape of a butterfly that covers a finite fraction the total field space. Imposing further 
the CMB constraints on the spectral 
index and the tensor ratio leads to initial values for which the spectral index admits an approximate scaling behavior that 
is reminiscent of the scaling behavior of monomial and Starobinsky inflation. Nevertheless the bounds on $r$ that have been discussed
in the literature in the context of singlefield inflation are not satisfied in $j$-inflation.
  A detailed analysis of the tensor ratio on the space of initial values instead shows 
  that upcoming gravity wave experiments that can reach down to $r\cong 10^{-4}$ and even smaller
 will not be able to distinguish between large and small  field inflation. These experiments are nevertheless important even in 
 the absence of a discovery because they provide essential constraints for model building. They will in particular constrain
 the parameter space for $j-$inflation, which presents a prime target for future observations.
 
The CMB constraints that have been obtained since COBE have established the amplitude of the scalar power spectrum, 
leading to a normalization of the overall energy scale of inflationary models. In $j$-inflation as described in the present 
paper this leaves the second energy parameter $\mu$, which is constrained by the phenomenological parameters just discussed. 
This energy scale uniquely determines the curvature scalar of the field space, hence these observational constraints 
put bounds on how strongly curved this target space can be. 

\vskip .3truein

{\large \bf Acknowledgement.} \\
It is a pleasure to thank Monika Lynker and Deepu Sengupta for discussions. This work was supported by 
a sabbatical leave of absence from Indiana University South Bend.

\end{document}